\documentclass[12pt]{article}

\usepackage{amsmath}
\usepackage{graphicx,psfrag,epsf}
\usepackage{enumerate}
\usepackage{graphicx}
\usepackage{amsmath, amsfonts, amssymb, mathrsfs, rotating, setspace}
\usepackage{textcomp}
\usepackage{enumitem}
\usepackage[utf8]{inputenc}
\usepackage[margin=1in]{geometry}
\usepackage{url} 
\usepackage{algorithm,algorithmic}
\usepackage{natbib}
\usepackage{xcolor}
\usepackage{bbm}
\usepackage{rotating}
\usepackage{booktabs}
\usepackage{caption}
\usepackage{subcaption}
\usepackage{hyperref}
\usepackage{url}
\usepackage{natbib}
\usepackage{breakurl}
\usepackage{tablefootnote}
\usepackage{booktabs,caption}
\usepackage{tikz}
\usepackage{soul}

\usetikzlibrary{calc}
\usetikzlibrary{positioning}
\usetikzlibrary{arrows}

\usepackage[flushleft]{threeparttable}

\newcommand{\blind}{1}

\begin{document}

\def\spacingset#1{\renewcommand{\baselinestretch}%
{#1}\small\normalsize} \spacingset{1}


\if1\blind
{
  \title{\bf Causal bounds for outcome-dependent sampling in observational studies}
  \author{Erin E. Gabriel$^{1\star}$\thanks{EEG is partially supported by Swedish Research Council grant 2017-01898, AS by Swedish Research Council grant 2016-01267 and MCS by Swedish Research Council grant 2019-00227} \hspace{.2cm}\\
  Michael C. Sachs$^1$\\
Arvid Sj\"olander$^{1\star}$\\
\footnotesize 1:Department of Medical Epidemiology and Biostatistics, Karolinska Institutet, Stockholm, Sweden \\
$\star$The authors contributed equally to this work.\\
corresponding author: erin.gabriel@ki.se}
  \maketitle
} \fi 

\if0\blind
{
  \bigskip
  \bigskip
  \bigskip
  \begin{center}
    {\LARGE\bf Causal bounds for outcome-dependent sampling in observational studies}
\end{center}
  \medskip
} \fi

\bigskip
\begin{abstract}
Outcome-dependent sampling designs are common in many different scientific fields including epidemiology, ecology, and economics. As with all observational studies, such designs often suffer from unmeasured confounding, which generally precludes the nonparametric identification of causal effects. Nonparametric bounds can provide a way to narrow the range of possible values for a nonidentifiable causal effect without making additional untestable assumptions. The nonparametric bounds literature has almost exclusively focused on settings with random sampling, and the bounds have often been derived with a particular linear programming method. We derive novel bounds for the causal risk difference, often referred to as the average treatment effect, in six settings with outcome-dependent sampling and unmeasured confounding for a binary outcome and exposure. Our derivations of the bounds illustrate two approaches that may be applicable in other settings where the bounding problem cannot be directly stated as a system of linear constraints. We illustrate our derived bounds in a real data example involving the effect of vitamin D concentration on mortality. 
\end{abstract}

\noindent%
\vfill

\newpage
\spacingset{2}
\vspace{-1cm}
\section{Introduction}
The aim of empirical research is often to estimate the causal effect of a particular exposure on a particular outcome. In observational (i.e. not randomized) studies, this is typically complicated by the fact that there are confounders (i.e. common causes) of the exposure and the outcome. Often, these confounders are at least partially unmeasured, in which case the causal effect of interest is generally nonidentifiable as any observable association could be due to the uncontrolled common causes. 

One way to narrow the possible range of a nonidentifiable causal effect is to derive bounds, i.e. a range of values that are guaranteed to contain the true causal effect given the true distribution of the observed data.   \citep[e.g.][]{robins1989analysis,balke97,zhang2003estimation,cai2008bounds,sjolander2009bounds}. Generally, the width of these bounds depends on the amount of information provided by the study design; the more information the tighter the bounds. In particular, \citet{balke97} showed that the bounds for the causal risk difference may be substantially improved upon if an instrumental variable (IV) is available. In their work, \citet{balke1994counterfactual} developed a method for deriving valid and tight bounds when the causal effect of interest and the constraints implied by the causal model can be stated as a linear optimization problem. Much of the subsequent work on nonparametric causal bounds has used this method. 

The literature on nonparametric bounds for causal effects has almost exclusively focused on simple random sampling; \citet{kuroki2010sharp} is a notable exception. However, in many studies the probability of selection into the study depends on the outcome of interest. Such outcome-dependent sampling often has advantages over simple random sampling, such as increased statistical power when the outcome is rare, but it also complicates the analysis and interpretation of results \citep{didelez10}. The case-control design, and its variations, is perhaps the most common outcome-dependent design. Ideally, in this design, the sampling only depends on the outcome, and not on other variables in the study; we refer to this and all similar studies as \emph{uncounfounded outcome-dependent sampling}. It is in this ideal setting that the bounds derived in \citet{kuroki2010sharp} apply. However, this ideal may often be difficult to achieve in reality, as cases are often selected, at least partially, due to diagnosis.  

More often, selection will be confounded with the outcome and exposure. A test-negative design is an example of a potentially confounded outcome-dependent sampling design. In a test-negative study, subjects are selected from the group of patients seeking health care for a particular set of symptoms that are indicative of the disease of interest. Subjects are included in the study if they are tested for disease status based on their symptoms, and the exposure of interest is then retrospectively ascertained. This design, and variations of it, have been discussed extensively in the infectious disease literature \citep{sullivan15, sullivan16}. An important difference from the case-control design is that potentially unobserved characteristics, such as the subject's lifestyle or health-seeking behavior, may influence whether the subject is tested and therefore selected into the test-negative study. Although the ideal test-negative design will hold these unmeasured characteristics constant, if this is not the case and these characteristics are associated with the exposure (e.g. vaccination status), then the sampling-outcome relationship is confounded. We refer to this and all similar studies as \emph{confounded outcome-dependent sampling}.  

In addition to being confounded with the outcome and exposure, sampling may depend directly on the exposure. We refer to this as \emph{confounded exposure- and outcome-dependent sampling}. Although this is a setting that one hopes to avoid when sampling is outcome-dependent, it may arise in test-negative studies when receiving the exposure reduces one's threshold for seeking medical treatment. This may also be the case in studies where the exposure is itself a medical condition and having such a condition leads to increased medical monitoring and therefore an increased probability of being included in the study independent of the outcome. 

In this paper we derive bounds for the causal risk difference for a binary outcome and exposure. We consider all three sampling designs defined above, both when there is and is not an available IV. In the test-negative design, which is a confounded sampling, outcome-, and possibly exposure-dependent design, a possible IV is randomization of individuals at health care centers to be encouraged to take an intervention, for example a vaccine. Individuals would then be free to decide if they wished to use the vaccine, regardless of their randomized encouragement. For the unconfounded outcome-dependent sampling designs, a possible IV may be a genetic allele, as in so-called Mendelian randomization studies \citep{bowden2011mendelian}.    

Under outcome-dependent sampling we cannot use the linear programming method of \citet{balke1994counterfactual} directly, since conditioning on being selected into the study generally implies a nonlinear structure on the counterfactual probabilities. This is because conditional probabilities of the form $p\{D|S\}=p\{D,S\}/p\{S\}$, where $S$ stands for `selection' and $D$ is generic for `data', do not simplify to linear expressions in the counterfactuals even if $p\{D,S\}$ and $p\{S\}$ do \citep{kuroki2010sharp}. Furthermore, when the sampling is unconfounded, the lack of common causes of the outcome and the sampling implies certain independencies between the counterfactuals, which induces nonlinear constraints through the factorization of the joint distribution of these counterfactuals.  Therefore, we derive valid bounds in these nonlinear settings using approaches that may be applicable in other nonlinear settings. In addition, we numerically compare the bounds derived in each setting to each other and random sampling via simulations and in a real data example. This describes, in terms possibly easier to convey to practitioners, the information loss associated with outcome-dependent sample selection. 

The paper is structured as follows. In Section \ref{sec:not} we introduce basic notation and assumptions, and define the causal target parameter. In Section \ref{boundsprev} we outline the pertinent previously derived bounds. In Section \ref{bounds} we present our bounds for the target parameter and outline some possible approaches to deriving bounds in nonlinear settings. In Section \ref{refine} we make a qualitative comparison of the bounds, and discuss how the valid but not tight bounds can be improved. In Section \ref{est} we outline some possible ways to quantify sampling uncertainty when the bounds are estimated from observed data. In Sections \ref{num} and \ref{real} we make a quantitative comparison of the bounds through simulation and a real data example, respectively. We conclude and discuss directions for future work in Section \ref{dis}. R code for the simulation study and real data analysis is available as Supplementary Material, and on the author's Github \url{https://github.com/eegabriel/outdep_bounds}.

\section{Preliminaries} \label{sec:not}

\subsection{Notation and target parameter}
Let $X$ and $Y$ be the binary exposure and outcome of interest, and $Y(x)$ be the potential (or counterfactual) outcome for a given subject, if the exposure were set to level $x$ \citep{rubin1974estimating,pearl2009causality}, where $x, X \mbox{ and } Y(x) \in \{0, 1\}$. Let $Z \in \{0, 1\}$ be a valid and potentially available binary IV for $X$ and $Y$; we refer readers to \citet{glymour2012credible} for details on testing for validity of a candidate IV. Let $S$ be an indicator of being selected into the study; $S=1$ for ``selected'' and $S=0$ for ``not selected''. Let $U$ represent the full set of confounders for $X$, $Y$, and possibly $S$; there are no restrictions on the distribution of $U$, and we do not assume that it is observed. Thus, the observed data distribution is given by $p\{Z,X,Y|S=1\}$ when an IV is available, and by $p\{X,Y|S=1\}$ when an IV is not available; $p\{\cdot\}$ denotes the probability mass function. Because $U$ is unmeasured, no counterfactual probabilities  of the type $p\{Y(x)=y\}$ are identifiable in any of our settings of interest. Our target parameter is the causal risk difference
$$\theta=p\{Y(x=1)=1\}-p\{Y(x=0)=1\},$$ 
often referred to as the Average Treatment Effect (ATE). In terms of factual (in contrast to counterfactual) probability distributions, $\theta$ can be expressed as \begin{eqnarray*}
\theta&=&E[p\{Y(x=1)=1|U\}-p\{Y(x=0)=1|U\}]\\
&=&E[p\{Y(x=1)=1|X=1,U\}-p\{Y(x=0)=1|X=0,U\}]\\
&=&E[p\{Y=1|X=1,U\}-p\{Y=1|X=0,U\}],
\end{eqnarray*}
where the expectation is taken over the marginal distribution of $U$. The first equality follows from the law of total probability, the second equality follows from conditional exchangeability, given $U$, which means that $Y(x)$ and $X$ are conditionally independent, given $U$, and the third equality follows by consistency of counterfactuals, which means that $Y(x)=Y$ when $X=x$ \citep{rubin1974estimating,pearl2009causality}.  

We define the following short-hand notation: $p_{xy}= p\{X=x, Y=y\};  p_{xy.s}=p\{X=x, Y=y|S=s\}; p_{xys}= p\{X=x, Y=y, S=s\}; p_{xy.zs}=p\{X=x, Y=y|Z=z,S=s\}; p_{xys.z}=p\{X=x, Y=y, S=s|Z=z\}; r=p\{S=1\}; r_z=p\{S=1|Z=z\}$.

\subsection{Settings}
\begin{figure*}[h]
\captionsetup[sub]{width=.9\linewidth}
\centering
\resizebox{\linewidth}{!}{
\begin{subfigure}[t]{0.5\textwidth}
\centering
\begin{tikzpicture}
\node (uh) at (1,1) {$U$};
\node (v) at (0,0) {$X$};
\node (i) at (2,0) {$Y$};

\draw[-latex] (v) -- (i);

\draw[-latex] (uh) -- (v);
\draw[-latex] (uh) -- (i);


\end{tikzpicture}
\caption{Random sampling, without IV.\label{a}}
\end{subfigure}

\begin{subfigure}[t]{0.5\textwidth}
\centering
\begin{tikzpicture}
\node (E) at (-2,0) {$Z$};
\node (uh) at (1,1) {$U$};
\node (v) at (0,0) {$X$};
\node (i) at (2,0) {$Y$};
\draw[-latex] (E) -- (v);


\draw[-latex] (v) -- (i);

\draw[-latex] (uh) -- (i);

\draw[-latex] (uh) -- (v);


\end{tikzpicture}
\caption{Random sampling, with IV.\label{b}}
\end{subfigure}
}

\vspace{4mm}

\captionsetup[sub]{width=.9\linewidth}
\centering
\resizebox{\linewidth}{!}{
\begin{subfigure}[t]{0.5\textwidth}
\centering
\begin{tikzpicture}
\node (uh) at (1,1) {$U$};
\node (v) at (0,0) {$X$};
\node (i) at (2,0) {$Y$};
\node[draw, rectangle] (s) at (4,0) {$S$};
\draw[-latex] (v) -- (i);
\draw[-latex] (i) -- (s);

\draw[-latex] (uh) -- (v);
\draw[-latex] (uh) -- (i);


\end{tikzpicture}
\caption{Unconfounded outcome-dependent sampling, without IV.\label{c}}
\end{subfigure}

\begin{subfigure}[t]{0.5\textwidth}
\centering
\begin{tikzpicture}
\node (E) at (-2,0) {$Z$};
\node (uh) at (1,1) {$U$};
\node (v) at (0,0) {$X$};
\node (i) at (2,0) {$Y$};
\draw[-latex] (E) -- (v);

\node[draw, rectangle] (s) at (4,0) {$S$};

\draw[-latex] (v) -- (i);
\draw[-latex] (i) -- (s);
\draw[-latex] (uh) -- (i);

\draw[-latex] (uh) -- (v);


\end{tikzpicture}
\caption{Unconfounded outcome-dependent sampling, with IV.\label{d}}
\end{subfigure}
}

\vspace{4mm}

\captionsetup[sub]{width=.9\linewidth}
\centering
\resizebox{\linewidth}{!}{
\begin{subfigure}[t]{0.5\textwidth}
\centering
\begin{tikzpicture}
\node (uh) at (1,1) {$U$};
\node (v) at (0,0) {$X$};
\node (i) at (2,0) {$Y$};
\node[draw, rectangle] (s) at (4,0) {$S$};
\draw[-latex] (v) -- (i);
\draw[-latex] (i) -- (s);
\draw[-latex] (uh) -- (i);
\draw[-latex] (uh) -- (s);

\draw[-latex] (uh) -- (v);

\end{tikzpicture}
\caption{Confounded outcome-dependent sampling, without IV.\label{e}}
\end{subfigure}

\begin{subfigure}[t]{0.5\textwidth}
\centering
\begin{tikzpicture}
\node (E) at (-2,0) {$Z$};
\node (uh) at (1,1) {$U$};
\node (v) at (0,0) {$X$};
\node (i) at (2,0) {$Y$};
\draw[-latex] (E) -- (v);

\node[draw, rectangle] (s) at (4,0) {$S$};

\draw[-latex] (v) -- (i);
\draw[-latex] (i) -- (s);
\draw[-latex] (uh) -- (i);
\draw[-latex] (uh) -- (s);

\draw[-latex] (uh) -- (v);

\end{tikzpicture}
\caption{Confounded outcome-dependent sampling, with IV.\label{f}}
\end{subfigure}
}

\vspace{4mm}

\captionsetup[sub]{width=.9\linewidth}
\centering
\resizebox{\linewidth}{!}{
\begin{subfigure}[t]{0.5\textwidth}
\centering
\begin{tikzpicture}
\node (uh) at (1,1) {$U$};
\node (v) at (0,0) {$X$};
\node (i) at (2,0) {$Y$};
\node[draw, rectangle] (s) at (4,0) {$S$};
\draw[-latex] (v) -- (i);
\draw[-latex] (i) -- (s);
\draw[-latex] (uh) -- (i);
\draw[-latex] (uh) -- (s);

\draw[-latex] (uh) -- (v);
\draw[->] (v) to[out=-45,in=-135] (s);

\end{tikzpicture}
\caption{Confounded exposure- and outcome-dependent sampling, without IV.\label{g}}
\end{subfigure}

\begin{subfigure}[t]{0.5\textwidth}
\centering
\begin{tikzpicture}
\node (E) at (-2,0) {$Z$};
\node (uh) at (1,1) {$U$};
\node (v) at (0,0) {$X$};
\node (i) at (2,0) {$Y$};
\draw[-latex] (E) -- (v);

\node[draw, rectangle] (s) at (4,0) {$S$};

\draw[-latex] (v) -- (i);
\draw[-latex] (i) -- (s);
\draw[-latex] (uh) -- (i);
\draw[-latex] (uh) -- (s);
\draw[->] (v) to[out=-45,in=-135] (s);
\draw[-latex] (uh) -- (v);

\end{tikzpicture}
\caption{Confounded exposure- and outcome-dependent sampling, with IV.\label{h}}
\end{subfigure}
}

\caption{Causal diagrams for eight possible study designs; a and b represent random sampling and c-h represent outcome-dependent sampling. \label{DAGS}}
\end{figure*}
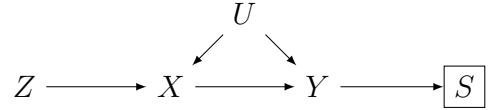
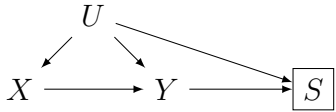
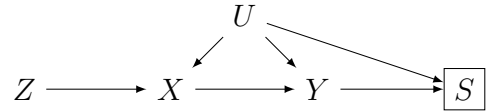
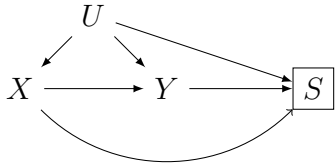
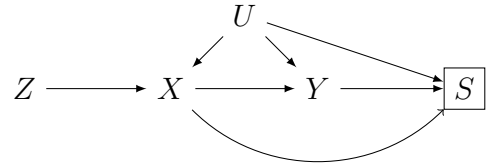

The causal diagrams in Figures \ref{a}, \ref{c}, \ref{e} and \ref{g} illustrate random, unconfounded outcome-dependent, confounded outcome-dependent, and confounded exposure- and outcome-dependent sampling, respectively, without an available IV \citep{pearl2009causality}. The causal diagrams in Figures \ref{b}, \ref{d}, \ref{f} and \ref{h} illustrate the corresponding sampling designs with an available IV. The square around $S$ in Figures \ref{c}, \ref{d}, \ref{e}, \ref{f}, \ref{g} and \ref{h} indicates that the observed data distribution is conditioned on $S=1$, i.e. on selection into the study. The presence of an arrow from $Y$ to $S$ in the figures makes the sampling outcome-dependent. The absence of an arrow from $U$ to $S$ in Figures \ref{c} and \ref{d} makes the sampling unconfounded, whereas the presence of this arrow in Figures \ref{e}-\ref{h} makes the sampling confounded. In Figures \ref{b}, \ref{d}, \ref{f} and \ref{h}, $Z$ does not have a direct effect on $Y$, and $Z$ has a causal effect on $X$. This is stronger than is necessary; in order for $Z$ to be a valid IV it is enough for $Z$ to be associated with $X$, and only have an effect on $Y$ via $X$. We assume throughout that it is known based on prior subject matter knowledge that one of the causal diagrams in Figures \ref{a}-\ref{h} fits the data at hand.  

\subsection{Available information} 
\label{sec:available}
To provide bounds on $\theta$ under outcome-dependent sampling, it is useful to have knowledge of quantities that are not given by the observed data distribution. Throughout, we will assume that the selection probability $p\{S=1\}$ is known, or can be estimated. 

The probability $p\{S=1\}$ should be thought of as reflecting the underlying mechanism through which subjects in the super-population become eligible for being sampled into the study. In practice, the selected and observed sample of size $n$ is usually taken from a finite sampling frame or cohort with known size $N$,  which is itself a random sample from a super-population. It is then natural to equate $p\{S=1\}$ with $n/N$, which may be viewed as a fixed constant or a random estimator, depending on the study design. 

For instance, under unconfounded outcome-dependent sampling where one draws a fixed number of cases, $n_1$, and controls, $n_0$, from a finite random sample of size $N$, the ratio $n/N=(n_1+n_0)/N=p\{S=1\}$ is a fixed constant by design. Alternatively, if the conditional selection probabilities $p\{S = 1|Y = 1\}$ and $p\{S = 1|Y = 0\}$ are fixed and the outcome prevalence in the super-population, $p\{Y = 1\}$, can be considered known, then $p\{S=1\}=p\{S = 1|Y = 1\}p\{Y=1\}+p\{S = 1|Y = 0\}p\{Y=0\}$ would also be considered fixed and known by design. 

Under other forms of outcome-dependent sampling, $p\{S=1\}$ may not be known or fixed by design. For example, under the test-negative design, a number of $n$ subjects are tested for disease status out of a random sample of fixed size $N$. If $p\{S=1\}$ is not known in this setting, one can still estimate it by  $n/N$. However, as $n$ is not fixed by design, this is a random estimator. Whether $p\{S=1\}$ is known and fixed or estimated and random has consequences when accounting for uncertainty in the estimation of the bounds; we return to this issue in Section \ref{est}.

When there is an available IV (Figures \ref{d}, \ref{f} and \ref{h}) we additionally assume that the IV prevalence in the super-population, $p\{Z=1\}$, can be considered known. When the IV is the randomization of individuals in health care centers to provide the vaccine, as may be the case in a test-negative study, $p\{Z=1\}$ is determined by design. When the IV is a genetic allele, $p\{Z=1\}$ is often known with high accuracy from previous research. We use the observed data distribution together with $p\{S=1\}$ and $p\{Z=1\}$ to calculate $p\{S=1|Z=z\}$, for $z \in \{0,1\}$, as $p\{S=1|Z=z\}=p\{Z=z|S=1\}p\{S=1\}/p\{Z=z\}$. 

When the sampling is outcome-dependent but unconfounded, $X$ and $Z$ are independent of $S$ given $Y$ (Figures \ref{c} and \ref{d}). Therefore, if $p\{Y\}$ were known, one could recover the unconditional distributions $p\{X,Y\}$ and $p\{Z,X,Y\}$ from the observed conditional distributions $p\{Z,X,Y|S=1\}$ and $p\{X,Y|S=1\}$, and apply previously developed bounds for $\theta$ under random sampling. However, knowing the outcome prevalence is not sufficient to recover the  unconditional distributions $p\{X,Y\}$ and $p\{Z,X,Y\}$ when the outcome-dependent sampling is confounded (Figures \ref{e}-\ref{h}). Therefore, to allow for equitable comparison of the derived bounds, we consider the same external information in all scenarios, i.e. $p\{S=1\}$, without an IV and $p\{S=1\}$ and $p\{Z=1\}$ with an IV. 

\section{Previous bounds} \label{boundsprev}
\citet{robins1989analysis} derived bounds for $\theta$ under random sampling. Without an available IV (Figure \ref{a}), these bounds are given by 
\begin{eqnarray}
\label{Beq:1}
 -(p_{01}+p_{10}) \leq \theta \leq 1-(p_{01}+p_{10}).
\end{eqnarray}
\citet{robins1989analysis} also derived bounds in the setting when an IV is available (Figure \ref{b}). However, \citet{balke97} showed that those bounds are not tight, using the linear programming method of bounds derivation developed in \citet{balke1994counterfactual}. Briefly, this method uses the fact that if all of the observed data probabilities conditional on the IV can be written as linear combinations of underlying counterfactual probabilities, then maximizing and minimizing the causal risk difference, $\theta$, under the linear constraints defined by these relations is a linear programming problem.

Since the IV is unconfounded with the outcome, it can be shown that optimization under those constraints on the conditional probabilities yields global extrema and therefore tight bounds on $\theta$. ``Tight'' here means that there are no values inside the bounds that are not logically compatible with the observed data distribution, under the assumptions (e.g. causal diagram) that were used to derive the bounds. These extrema can be obtained symbolically using fundamental concepts in the field of linear programming, specifically vertex enumeration \citep{DantzigLP}. The bounds derived by \citep{balke97} with an available IV have been cited numerous times in the causal inference literature \citep[e.g.][]{greenland2000introduction,hernan2006instruments}; we reproduce these bounds in Equations (1) and (2) of the Supplementary Material. 

\citet{kuroki2010sharp} derived bounds conditional on $Y$, which are applicable to the settings in Figure \ref{a} and \ref{c}; we refer to their bounds as ``outcome-conditional bounds''. As noted by \citet{kuroki2010sharp}, their outcome-conditional bounds are generally wider than the bounds in (\ref{Beq:1}), since they do not use the information captured by the marginal distribution of $Y$.

\section{Novel Bounds} \label{bounds}
\subsection{Unconfounded outcome-dependent sampling}
Unconfounded outcome-dependent sampling is illustrated in Figures \ref{c} and \ref{d}, without and with an available IV, respectively. In these settings we cannot use the linear programming method of \citet{balke1994counterfactual} directly, because conditioning on $S=1$ implies a nonlinear structure on the counterfactual probabilities and the absence of an arrow from $U$ to $S$ induces nonlinear constraints. Our approach to deriving bounds in these nonlinear settings has some generality, and may be used in other settings with nonlinear constraints due to an unconfounded non-outcome variable (e.g. $S$) and unobserved data for some levels of this variable (e.g. for $S=0$). 

Define $A(y,s)=p_{1y.s}/p_{0y.s}$. It can be shown (see the Supplementary Material) that the absent arrow from $U$ to $S$ in Figure \ref{c} implies the nonlinear constraint
\begin{eqnarray}
\label{eq:Ay}
A(y,0)=A(y,1)\textrm{ for all }y,
\end{eqnarray}
provided that sampling was not deterministic, i.e. $0<p\{S=1|Y=y\}<1$, $\forall y$. A similar nonlinear constraint would hold in other settings with an unconfounded non-outcome variable and partially unobserved data. To derive bounds that take the nonlinear constraint into account we proceed as follows. We start with valid and tight bounds under a setting where the unconfounded non-outcome variable is absent and data are fully observed, e.g. the bounds in (\ref{Beq:1}) corresponding to the causal diagram in Figure \ref{a}. We then partition those bounds into observed and unobserved parts, by conditioning on the unconfounded variable. Finally, we use the nonlinear constraint to bound the unobserved part, thus producing bounds for the target parameter. For example, we partition the term $p_{01}+p_{10}$ in \eqref{Beq:1} as
\begin{eqnarray}
\label{eq:partitioning}
p_{01}+p_{10}=(p_{01.1}+p_{10.1})r+(p_{01.0}+p_{10.0})(1-r).
\end{eqnarray}
We show in the Supplementary Material that, under the nonlinear constraint in \eqref{eq:Ay}, the unobserved part $p_{01.0}+p_{10.0}$ is bounded by 
\begin{eqnarray}
\label{eq:arvid1}
\min\left\{\frac{1}{1+A(1,1)},\frac{A(0,1)}{1+A(0,1)}\right\}\leq p_{01.0}+p_{10.0}\leq\max\left\{\frac{1}{1+A(1,1)},\frac{A(0,1)}{1+A(0,1)}\right\}.
\end{eqnarray}
Similar bounds can be obtained under other nonlinear constraint in other settings, by following the arguments in the Supplementary Material. Combining \eqref{eq:partitioning} and \eqref{eq:arvid1} with \eqref{Beq:1} gives the bounds for $\theta$ in (\ref{Beq:5}) and (\ref{Beq:6}).  

Analogously, define $B(y,z,s)=p_{1y.zs}/p_{0y.zs}$. It can be shown (see the Supplementary Material) that the absent arrow from $U$ to $S$ in Figure \ref{d} implies the nonlinear constraint
\begin{eqnarray}
\label{eq:BBy}
B(y,z,0)=B(y,z,1)\textrm{ for all }(y,z),
\end{eqnarray}
provided that sampling was not deterministic. To derive bounds that take this nonlinear constraint into account we start with the valid and tight bounds under Figure \ref{b}, derived by \citet{balke97}, where $S$ is absent and data are fully observed. Partitioning these bounds into observed and unobserved parts, similar to (\ref{eq:partitioning}), and using the nonlinear constraint in (\ref{eq:BBy}) to bound the unobserved parts, similar to (\ref{eq:arvid1}), gives the bounds in \eqref{Beq:7} and \eqref{Beq:8}.

\noindent \textbf{Result 1}:\\
The bounds given in \eqref{Beq:5} and \eqref{Beq:6} are valid for $\theta$ in the setting of Figure \ref{c} provided that $p_{0y.1} \neq 0$ $\forall y$, or for $\theta^*=-\theta$ if $x$ is replaced with $x^* = 1 - x$ everywhere in \eqref{Beq:5} and \eqref{Beq:6}, provided that $p_{1y.1} \neq 0$ $\forall y$. 
\begin{eqnarray}
\label{Beq:5}
\theta\geq -(p_{01.1}+p_{10.1})r-\max\left\{\frac{1}{1+A(1,1)},\frac{A(0,1)}{1+A(0,1)}\right\}(1-r)
\end{eqnarray}
and
\begin{eqnarray}
\label{Beq:6}
\theta\leq 1-(p_{01.1}+p_{10.1})r-\min\left\{\frac{1}{1+A(1,1)},\frac{A(0,1)}{1+A(0,1)}\right\}(1-r).
\end{eqnarray}

\noindent \textbf{Result 2}:\\
The bounds given in \eqref{Beq:7} and \eqref{Beq:8} are valid for $\theta$ in the setting of Figure \ref{d} provided  $p_{0y1.z} \neq 0$ $\forall y,z$, or for $\theta^*=-\theta$ if $x$ is replaced everywhere in \eqref{Beq:7} and \eqref{Beq:8} with $x^* = 1 - x$, provided that  $p_{1y1.z} \neq 0$ $\forall y,z$. 

Detailed derivations of Results 1 and 2 are given in the Supplementary Materials. We note that the bounds in Result 1 are only valid and informative if $A(y,1)$ is defined; the ratio $A(y,1)$ is undefined if $p_{0y.1}=0$. A solution to this problem is to reverse the coding for the exposure, i.e. to define a new exposure as $x^*=1-x$ for which $A(y,1)$ are inverted. Replacing $x$ everywhere in \eqref{Beq:5} and \eqref{Beq:6} with $x^*$ we can obtain the lower bound, $l^*$, and upper bound, $u^*$, for $\theta^*=p\{Y(x^*=1)=1\}-p\{Y(x^*=0)=1\}$. Since $\theta^*=-\theta$, these translate into bounds for $\theta$ as $-u^*\leq\theta\leq -l^*$. 

\begin{eqnarray}
\label{Beq:7}
\theta\geq \max\left\{\begin{array}{l}p_{11.11}r_1+p_{00.01}r_0-1\\
p_{11.01}r_0+p_{00.11}r_1-1\\
(p_{11.01}-p_{10.01}-p_{01.01})r_0-(p_{11.11}+p_{01.11})r_1-\frac{B(0,0,1)}{1+B(0,0,1)}(1-r_0)-(1-r_1)\\
(p_{11.11}-p_{10.11}-p_{01.11})r_1-(p_{11.01}+p_{01.01})r_0-\frac{B(0,1,1)}{1+B(0,1,1)}(1-r_1)-(1-r_0)\\
-(p_{10.11}+p_{01.11})r_1-\max\left\{\frac{1}{1+B(1,1,1)},\frac{B(0,1,1)}{1+B(0,1,1)}\right\}(1-r_1)\\
 -(p_{10.01}+p_{01.01})r_0-\max\left\{\frac{1}{1+B(1,0,1)},\frac{B(0,0,1)}{1+B(0,0,1)}\right\}(1-r_0)\\
 (p_{00.11}-p_{10.11}-p_{01.11})r_1-(p_{10.01}+p_{00.01})r_0\\
\phantom{11111}-\max\left\{\frac{1}{1+B(1,1,1)},-\frac{1-B(0,1,1)}{1+B(0,1,1)}\right\}(1-r_1)-(1-r_0)\\
(p_{00.01}-p_{10.01}-p_{01.01})r_0-(p_{10.11}+p_{00.11})r_1\\
\phantom{11111}-\max\left\{\frac{1}{1+B(1,0,1)},-\frac{1-B(0,0,1)}{1+B(0,0,1)}\right\}(1-r_0)-(1-r_1)
\end{array}\right\}, 
\end{eqnarray}
and 
\begin{eqnarray}
\label{Beq:8}
\theta\leq \min\left\{\begin{array}{l}1-p_{10.11}r_1-p_{01.01}r_0\\
1-p_{10.01}r_0-p_{01.11}r_1\\
(p_{10.11}+p_{00.11})r_1+(p_{11.01}+p_{00.01}-p_{10.01})r_0\\
\phantom{11111}+(1-r_1)+\max\left\{\frac{1-B(0,0,1)}{1+B(0,0,1)},\frac{B(1,0,1)}{1+B(1,0,1)}\right\}r_0\\
(p_{11.11}+p_{00.11}-p_{10.11})r_1+(p_{10.01}+p_{00.01})r_0\\
\phantom{11111}+\max\left\{\frac{1-B(0,1,1)}{1+B(0,1,1)},\frac{B(1,1,1)}{B(1,1,1)}\right\}(1-r_1)+(1-r_0)\\
(p_{11.11}+p_{00.11})r_1+\max\left\{\frac{1}{1+B(0,1,1)},\frac{B(1,1,1)}{1+B(1,1,1)}\right\}(1-r_1)\\
(p_{11.01}+p_{00.01})r_0+\max\left\{\frac{1}{1+B(0,0,1)},\frac{B(1,0,1)}{1+B(1,0,1)}\right\}(1-r_0)\\
(p_{11.11}+p_{00.11}-p_{01.11})r_1+(p_{11.01}+p_{01.01})r_0\\
\phantom{11111}+\max\left\{\frac{1}{B(0,1,1)},-\frac{1-B(1,1,1)}{1+B(1,1,1)}\right\}(1-r_1)+(1-r_0)\\
(p_{11.01}+p_{00.01}-p_{01.01})r_0+(p_{11.11}+p_{01.11})r_1\\
\phantom{11111}+\max\left\{\frac{1}{B(0,0,1)},-\frac{1-B(1,0,1)}{1+B(1,0,1)}\right\}(1-r_0)+(1-r_1)
\end{array}\right\}.
\end{eqnarray}

This simple solution works for the bounds if the inverse ratios, $A^{-1}(y,1)$, are not undefined. However, this may not be the case. It is possible that $p_{1y.1}=p_{0y.1}=0$ for some $y$, so that both $A(y,1)$ and $A^{-1}(y,1)$, are undefined. For such scenarios, we suggest using the bounds in \eqref{Beq:9}, for confounded sampling. Similarly for Result 2, when $B(y,z,1)$ is undefined for any $y,z$ pair, we can instead bound $-\theta$ by defining exposure as $x^*=1-x$, provided that $B^{-1}(y,z,1)$ is not undefined. When $B^{-1}(y,z,1)$ and $B(y,z,1)$ are both undefined for a given $y$ and all $z$, then we suggest using the bounds given in \eqref{Beq:11} and \eqref{Beq:12}, for confounded sampling in addition to any remaining defined terms from \eqref{Beq:7} and \eqref{Beq:8}.

\subsection{Confounded outcome-dependent sampling} \label{CCB}
Confounded outcome-dependent sampling is illustrated in Figures \ref{e} and \ref{f}, without and with an available IV, respectively. Confounded exposure- and outcome-dependent sampling is illustrated in Figures \ref{g} and \ref{h}, without and with an available IV, respectively. In these settings, the arrow from $U$ to $S$ is present; thus, there are no nonlinear constraints on $A(y,s)$ or $B(y,z,s)$. However, the conditioning on $S=1$ still implies a nonlinear structure on the counterfactual probabilities. 

Under Figures \ref{e} and \ref{g} the observed data distribution is given by $p\{X,Y|S=1\}$, and under Figures \ref{f} and \ref{h} the observed data distribution is given by $p\{Z,X,Y|S=1\}$. The linear programming method of \citet{balke1994counterfactual} cannot be directly applied to these probabilities, since they are conditioned on $S=1$, and are therefore nonlinear functions of the counterfactual probabilities. However,  if we know $p\{S=1\}$ (or can derive it from known $p\{S=1|Y\}$ and $p\{Y\}$), then the joint probabilities $p\{X=x,Y=y,S=1\}=p\{X=x,Y=y|S=1\}p\{S=1\}$ are known or estimable. Similarly, if we know $p\{S=1\}$ and $p\{Z=1\}$, and therefore $p\{S=1|Z\}$, then the probabilities  $$p\{X=x,Y=y,S=1|Z=z\}=\frac{p\{Z=z,X=x,Y=y|S=1\}}{\sum_{x,y}p\{Z=z,X=x,Y=y|S=1\}}p\{S=1|Z=z\}$$ are known or estimable. Arguing as in \citet{balke97}, the probabilities $p\{X=x,Y=y,S=1\}$ and $p\{X=x,Y=y,S=1|Z=z\}$ can be expressed as linear functions of the counterfactual probabilities. Therefore, knowing $p\{S=1\}$ and (for \ref{f} and \ref{h}) $p\{Z=1\}$ makes the problem of bounding $\theta$ a linear programming problem, to which the method of \citet{balke1994counterfactual} can be applied. Details are available in the Supplementary Material.

\noindent \textbf{Result 3}: 
The bounds for $\theta$ given in \eqref{Beq:9} are valid and tight in the setting of Figure \ref{e}, and the bounds given in \eqref{Beq:11} and \eqref{Beq:12} are valid and tight in the settings of Figure \ref{f}, provided that $p\{S=1\}$ and (for Figure \ref{f}) $p\{Z=1\}$ are known.
\vspace{-0.5cm}
\begin{eqnarray}
\label{Beq:9}
 p_{111}+p_{001}-1 \leq \theta \leq 1-p_{011}-p_{101}.
\end{eqnarray}

\noindent \textbf{Result 4}: 
The bounds for $\theta$ given in \eqref{Beq:9}, and in \eqref{Beq:13} and \eqref{Beq:14} are valid and tight in the settings of Figure \ref{g} and Figure \ref{h}, respectively, provided that $p\{S=1\}$ and (for Figure \ref{h}) $p\{Z=1\}$ are known.

\begin{eqnarray}
\label{Beq:11}
\theta\geq \max\left\{\begin{array}{l}
p_{001.1} - p_{011.0} - p_{111.0} + 2p_{111.1} - 1\\
p_{001.1} + p_{111.0} - 1\\
p_{001.0} + p_{111.1} - 1\\
p_{001.1} + p_{111.1} - 1\\
p_{001.0} - p_{011.1} + 2p_{111.0} - p_{111.1} - 1\\
- p_{001.0} + 2p_{001.1} - p_{101.0} + p_{111.1} - 1\\
p_{001.0} + p_{111.0} - 1\\
2p_{001.0} - p_{001.1} - p_{101.1} + p_{111.0} - 1\\
2p_{001.0} - p_{001.1} - p_{101.1} - p_{011.1} + 2p_{111.0} - p_{111.1} - 1\\
- p_{001.0} + 2p_{001.1} - p_{101.0} - p_{011.0} - p_{111.0} + 2p_{111.1} - 1
\end{array}\right\}.
\end{eqnarray}

\begin{eqnarray}
\label{Beq:12}
\theta\leq \min\left\{\begin{array}{l}
- p_{101.0} - 2p_{011.0} + p_{011.1} + p_{111.1} + 1\\
- p_{101.1} + p_{011.0} - 2p_{011.1} + p_{111.0} + 1\\
p_{001.1} - 2p_{101.0} + p_{101.1} - p_{011.0} + 1\\
- p_{101.0} - p_{011.1} + 1\\
- p_{101.1} - p_{011.0} + 1\\
- p_{101.0} - p_{011.0} + 1\\
p_{001.1} - 2p_{101.0} + p_{101.1} - 2p_{011.0} + p_{011.1} + p_{111.1} + 1\\
- p_{101.1} - p_{011.1} + 1\\
p_{001.0} + p_{101.0} - 2p_{101.1} - p_{011.1} + 1\\
p_{001.0} + p_{101.0} - 2p_{101.1} + p_{011.0} - 2p_{011.1} + p_{111.0} + 1\\
\end{array}\right\}.
\end{eqnarray}

\begin{eqnarray}
\label{Beq:13}
\theta\geq \max\left\{\begin{array}{l}
  p_{001.1} + p_{111.1}-1\\
  p_{001.0} + p_{111.1}-1\\
  p_{001.1} + p_{111.0}-1\\
  p_{001.0} + p_{111.0}-1\\
  2p_{001.1} + p_{011.0} + p_{111.0} + p_{111.1}-2\\
  2p_{001.0} + p_{011.1} + p_{111.0} + p_{111.1}-2\\
  p_{001.0} + p_{001.1} + p_{101.0} + 2p_{111.1}-2\\
  p_{001.0} + p_{001.1} + p_{101.1} + 2p_{111.0}-2
\end{array}\right\},
\end{eqnarray}
and
\begin{eqnarray}
\label{Beq:14}
\theta\leq \min\left\{\begin{array}{l}
   -p_{101.0}-p_{011.0} + 1\\
 -p_{101.1}-p_{011.0} + 1\\
 -p_{101.0}-p_{011.1} + 1\\
 -p_{101.1}-p_{011.1} + 1\\
 -p_{001.0}-p_{101.0}-p_{101.1}-2p_{011.1} + 2\\
 -p_{001.1}-p_{101.0}-p_{101.1}-2p_{011.0} + 2\\
 -2p_{101.0}-p_{011.0}-p_{011.1}-p_{111.1} + 2\\
 -2p_{101.1}-p_{011.0}-p_{011.1}-p_{111.0} + 2
\end{array}\right\}.
\end{eqnarray}

\section{Qualitative comparison and refinement of the bounds}  \label{refine}

The following result, which we prove in the Supplementary Material, shows the relation between our bounds in \eqref{Beq:5} and \eqref{Beq:6} and the outcome-conditional bounds of \citet{kuroki2010sharp}.

\noindent \textbf{Result 5}: The lower bound in \eqref{Beq:5} is monotonically increasing in $p\{S = 1\} = r$, and the upper bound in \eqref{Beq:6} is monotonically decreasing in $r$. The bounds in \eqref{Beq:5} and \eqref{Beq:6} converge to the outcome-conditional bounds of \citet{kuroki2010sharp} as $r$ goes to 0.

The result is intuitively reasonable; the larger the selected subpopulation, the more information it contains about the whole population, and the tighter the bounds. In the unconfounded outcome-dependent setting (Figure \ref{c}), when you condition on the outcome, sampling can be ignored because it is independent of the exposure given the outcome. However, when information about sampling probabilities is available, it can be used to get narrower bounds.

One would expect that the bounds in \eqref{Beq:5} and \eqref{Beq:6} (corresponding to Figure \ref{c}) are at least as wide as the bounds in \eqref{Beq:1}, and that the bounds in \eqref{Beq:9} (corresponding to Figures \ref{e} and \ref{g}) are at least as wide as the bounds in \eqref{Beq:5} and \eqref{Beq:6}. Using the relations in \eqref{eq:partitioning} and \eqref{eq:arvid1}, it can easily be shown that this is indeed the case. 

However, a similar relation does not hold for the bounds corresponding to Figures \ref{d} and \ref{f}. To show this, we give an example where probabilities $p_{zxy.1}$ are generated under the causal diagram in Figure \ref{d}, such that  the bounds in \eqref{Beq:11} and \eqref{Beq:12} are narrower than those in \eqref{Beq:7} and \eqref{Beq:8}. First, note that, under the causal diagram in Figure \ref{d}, the joint distribution of $(U,Z,X,Y,S)$ factorizes into $p\{U,Z,X,Y,S\}=p\{U)p\{Z\}p\{X|U,Z\}p\{Y|U,X\}p\{S|Y\}$. Now, suppose that $U$ is binary and that $p\{U=1\}=0.9$,
$p\{Z=1\}=0.7$, and the conditional probabilities are given in Table \ref{jointprop1}.

\begin{table}[h]
\centering
\begin{tabular}{lcccc|lcccc}
\multicolumn{5}{c}{$p\{X=x|U=u, Z=z\}$} & \multicolumn{5}{|c}{$p\{Y=y|U=u, X=x\}$} \\
\hline
&\multicolumn{4}{c|}{$uz$} &  &\multicolumn{4}{c}{$uz$} \\
&$00$&$01$&$10$&$11$& &$00$&$01$&$10$&$11$ \\
X=1&0.3&0.3&0.9&0.1&Y=1&0.3&0.4&0.1&0.8\\
\hline
\end{tabular}
\caption{Conditional probabilities}
\label{jointprop1}
\end{table}

As $U$ is the only confounder for $X$ and $U$, we have $p\{Y(x)=1|U\}=p\{Y=1|X=x,U\}$, so that $\theta=\sum_u[p\{Y=1|X=1,U=u\}-p\{Y=1|X=0,U=u\}]p\{U=u\}=0.64$. By marginalizing over $U$, the joint distribution for ($Z,X,Y,S$) becomes as in Table \ref{jointprop}. With this joint distribution, the lower and upper bounds using \eqref{Beq:7} and \eqref{Beq:8} (Figure \ref{d}) are equal to $-0.1678$ and $0.8067$, respectively, and the lower and upper using \eqref{Beq:11} and \eqref{Beq:12} (Figure \ref{f}) are equal to $0.1382$ and $0.8176$, respectively. 

\begin{table}[h]
    \centering
    \begin{tabular}{lcccccccc}
\hline
&\multicolumn{8}{c}{$xys$}\\
   &$001$&$011$&$101$&$111$&$000$&$010$&$100$&$110$\\
   \hline
      $z=0$& 0.03510& 0.00180 &0.04860&0.03960&
      0.00390&0.00720&0.00540&0.15840\\
    \hline
    $z=1$& 0.49014&0.01428&0.02268&0.01176&0.05446&0.05712&0.00252&0.04704\\
    \hline
    \end{tabular}
    \caption{Joint probabilities $p\{Z=z,X=x,Y=y,S=s\}$.}
    \label{jointprop}
\end{table}

An important implication of this example is that the bounds in \eqref{Beq:7} and \eqref{Beq:8} are not tight. This is because, if they were, they would never be wider than any other bounds for $\theta$ in the setting of Figure \ref{d} based on the same information. In particular, they would never be wider than the bounds in \eqref{Beq:11} and \eqref{Beq:12}, since we assume the same available information and the causal diagram in Figure \ref{d} is a special case of the causal diagram in Figure \ref{f}, which implies that the bounds in \eqref{Beq:11} and \eqref{Beq:12} are also valid for Figure \ref{d}. 

This argument, however, suggests a simple way to improve the bounds in \eqref{Beq:7} and \eqref{Beq:8}, namely to replace them with the bounds in \eqref{Beq:11} and \eqref{Beq:12} whenever these are tighter. Formally, let $l_d$ and $u_d$ be the lower and upper bounds in \eqref{Beq:7} and \eqref{Beq:8}, and let $l_f$ and $u_f$ be the lower and upper bounds in \eqref{Beq:11} and \eqref{Beq:12}. We thus define new bounds for $\theta$ under the causal diagram in Figure \ref{d} that will be used instead of the bounds in \eqref{Beq:7} and \eqref{Beq:8} for the remainder of the paper: 
\begin{eqnarray}
\label{Beq:15}
\max(l_d,l_f) \leq \theta \leq \min(u_d,u_f).
\end{eqnarray}
  
\section{Estimation of the bounds} \label{est}
All proposed bounds are functions of probabilities $p\{Z,X,Y|S=1\}$ or $p\{X,Y|S=1\}$, which can be estimated by their sample analogues to produce estimated bounds. To account for the statistical uncertainty in the estimates we suggest nonparametric bootstrap, which we illustrate the use of in both the simulations and the real data example.   

Whether $p\{S=1\}$ is known or estimated, as discussed in Section \ref{sec:available}, has consequences for how the bootstrap samples should be taken. If $p\{S=1\}$ is fixed by design, then bootstrap sampling of the same sample size can be used; we refer to this bootstrap sampling as Type A. For instance, if $p\{S=1\}$ is equated with $n/N$, where $n$ and $N$ are fixed by design, then samples of size $n$ are taken with replacement from the observed sample with size $n$. 

We note that, although $p\{S=1\}$ may be fixed by design, it may not be fully known. This would, for instance, be the case if the conditional selection probabilities $p\{S=1|Y=1\}$ and $p\{S=1|Y=0\}$ are known, but the outcome prevalence $p\{Y=1\}$ is not. This uncertainty can be accounted for in a sensitivity analysis by estimating the bounds over a grid of plausible values for $p\{S=1\}$ (or $p\{Y = 1\}$), and bootstrapping the estimated bounds at each point in the grid.  

If the selection probability is an estimate from the observed data, then the bootstrap should reflect the uncertainty in this estimate. For instance, if $p\{S=1\}$ is estimated by $n/N$, where $n$ is random, then samples of size $N$ are taken with replacement from the finite sampling frame/cohort of size $N$ setting all variables not observed in the sample of size $n$ to missing. In this way the number of subjects for which $X$ and $Y$ (and possibly $Z$) are observed, say $n'$, and hence also $n'/N$, varies across bootstrap samples; we refer to this bootstrap sampling as Type B.

\section{Simulation}  \label{num}

In this section we compare the derived bounds and investigate their properties through simulations. We have divided the section into two parts. In the first part we focus on the true bounds, ignoring sampling variability arising from estimation in finite samples. In the second part we assess finite sample inference for estimated bounds.  

\subsection{True bounds}

In this simulation we generated probability distributions $p\{U,Z,X,Y,S\}$ under the causal diagram in Figure \ref{h}, from the parametric model
\begin{eqnarray}
\label{eq:simmod}
\left .
\begin{array}{rcl}
p\{U=1\} & \sim & \mbox{Unif}(0,1) \\
p\{Z=1\} & \sim & \mbox{Unif}(0,1) \\
p\{X=1|U,Z\} & = & \mbox{expit}(\alpha_1+\alpha_2U+\alpha_3Z+\alpha_4UZ)\\
p\{Y=1|U,X\} & = & \mbox{expit}(\beta_1+\beta_2U+\beta_3X+\beta_4UX)\\
p\{S=1|U,Y,X\} & = & \mbox{expit}(\gamma_1+\gamma_2Y +\gamma_3U+ \gamma_4X)\\
(\alpha_1,\alpha_2,\alpha_3,\alpha_4,\beta_1,\beta_2,\beta_3,\beta_4,\gamma_1,\gamma_2) & \sim & N(0,5^2)\\ 
\gamma_3 & \sim & N(0,\sigma_U^2)\\
\gamma_4 & \sim & N(0,\sigma_X^2)
\end{array}
\right \}
\end{eqnarray}
where $\mbox{expit}(x) = e^x / (1 + e^x)$. The causal parameter of interest $\theta=p\{Y(x=1)=1\}-p\{Y(x=0)=1\}$ is equal to $\sum_u[\mbox{expit}(\beta_1+\beta_2u+\beta_3+\beta_4u)-\mbox{expit}(\beta_1+\beta_2u)]p\{U=u\}$, the actual value of which varies with the probability of $p\{U=u\}$ and the $\beta$ values. In this model, $\sigma_U$ and $\sigma_X$ determine the degree of exposure dependence and confounding of the sampling; e.g. if $\sigma_U=0$, then the sampling is unconfounded, and if $\sigma_X=0$, then the sampling does not depend on the exposure.  We first generated 100,000 distributions $p\{U,Z,X,Y,S\}$ from the model in \eqref{eq:simmod}, with $\sigma_U=\sigma_X=0$, corresponding to Figures \ref{c} and \ref{d}. For each distribution we computed our proposed bounds corresponding to the assumed causal diagrams in Figures \ref{c}-\ref{h}, even if this is not how the data were generated. For comparison we also computed the previously derived bounds for random sampling, without \citep{robins1989analysis} and with an IV \citep{balke97}. These bounds use the distributions $p\{X,Y\}$ and $p\{Z,X,Y\}$, respectively, which do not condition on $S=1$. Thus, these bounds are not applicable in practice in outcome-dependent sampling designs. 

\begin{figure}[ht]
\captionsetup[sub]{width=.9\linewidth}
\centering
\resizebox{\linewidth}{!}{
\begin{subfigure}[t]{0.5\textwidth}

\begin{center}		
\includegraphics[width=.98\textwidth]{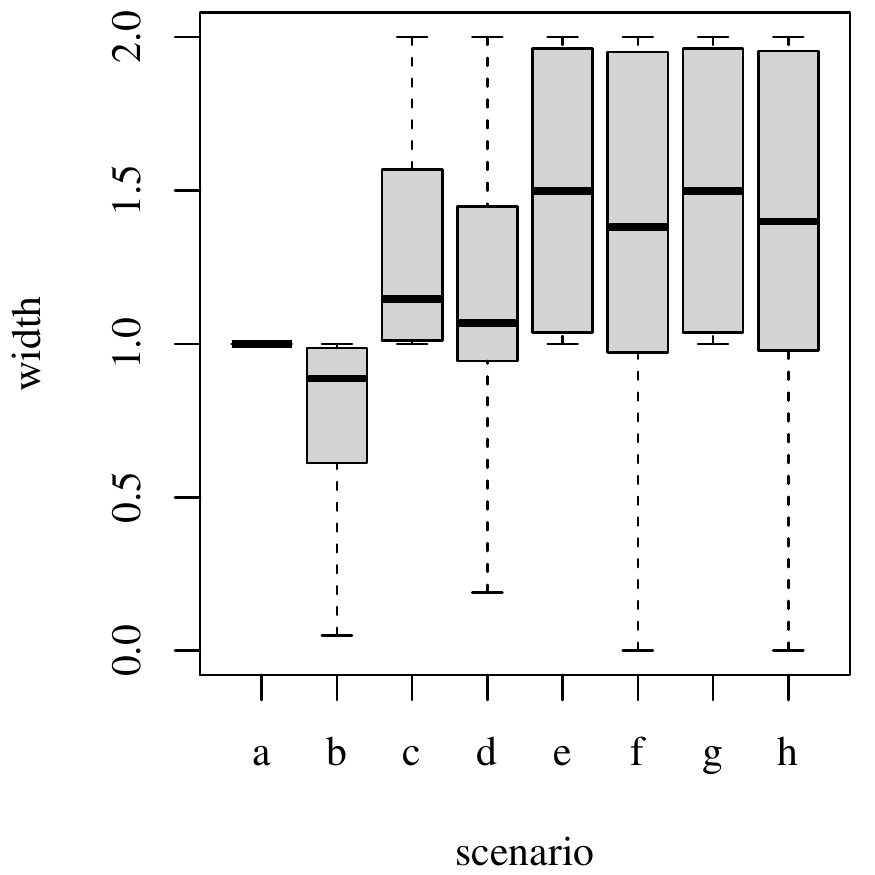}
\caption{Width of the bounds when $\sigma_U=\sigma_X=0$ (unconfounded outcome-dependent sampling).}
\label{fig:width0}
\end{center}
\end{subfigure}

\begin{subfigure}[t]{.5\textwidth}
\begin{center}		
\includegraphics[width=.98\textwidth]{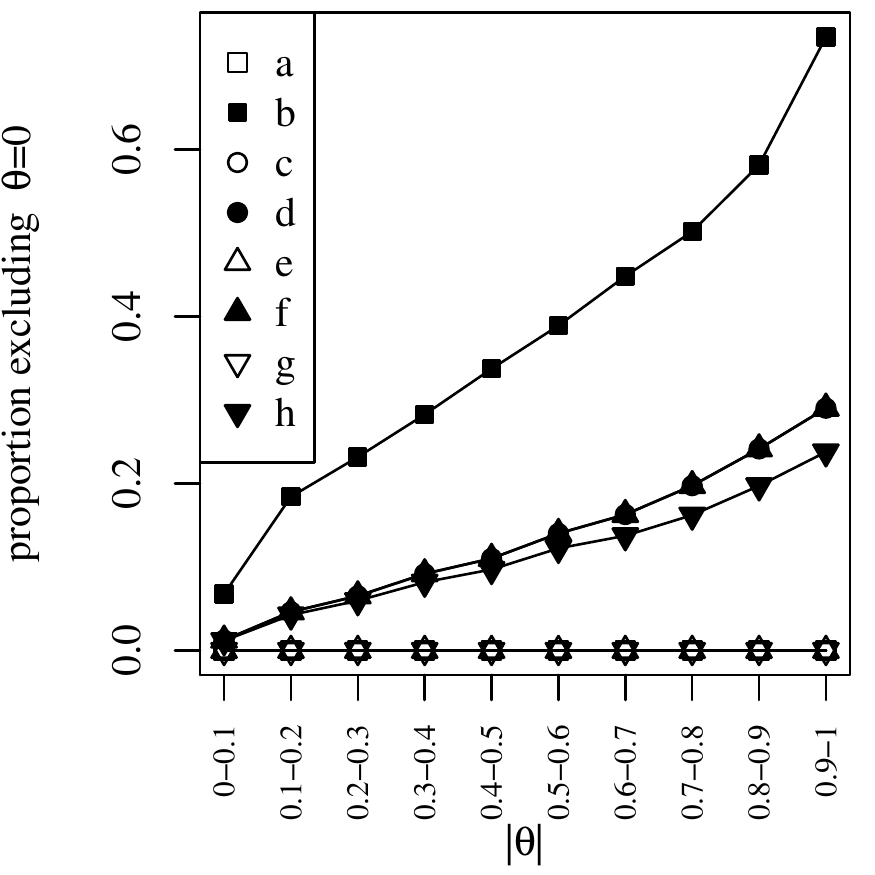}
\caption{Proportion of the bounds that exclude the causal null hypothesis $\theta=0$, as a function of $|\theta|$, when $\sigma_U=\sigma_X=0$ (unconfounded outcome-dependent sampling).}
\label{fig:exclude}
\end{center}
\end{subfigure}
}
\caption{Summary of simulation results.}
\end{figure}

Figure \ref{fig:width0} shows the distribution of the width of the bounds. As expected, we observe that an available IV generally makes the bounds narrower (b vs a, d vs c, f vs e, h vs g), and that stronger deviation from the ideal randomized trial with random sampling generally makes the bounds wider (a vs c vs e, b vs d vs f). However, an exception from the latter is when exposure dependence is allowed for; the bounds under Figures \ref{g} and \ref{e} are of course identical, and the bounds under Figures \ref{h} and \ref{f} are often very similar. With respect to the width of the bounds, the availability of an IV does not seem to compensate for a non-ideal sampling design; the median width under Figure \ref{d} (outcome-dependent unconfounded sampling, available IV) is larger than the median width under Figure \ref{a} (random sampling, no available IV), and the median width under Figure \ref{f} (outcome-dependent confounded sampling, available IV) is larger than the median width under Figure \ref{c} (outcome-dependent unconfounded sampling, no available IV).     

Apart from their width, an important property of the bounds is their ability to exclude the causal null hypothesis $\theta=0$, given that $\theta\neq 0$. In the simulation above, $\theta$ is a function of the confounder prevalence $p\{U=1\}$ and the parameters $(\beta_1,\beta_2,\beta_3,\beta_4)$ in the model for $p\{Y=1|U,X\}$, and thus varies across the 100,000 generated distributions. Figure \ref{fig:exclude} shows the proportion of the bounds that exclude the causal null hypothesis $\theta=0$, as a function of $|\theta|$. The exclusion proportions for the bounds under Figure \ref{d} are not clearly visible in Figure \ref{fig:exclude}, since they are identical to the exclusion proportions for the bounds under Figure \ref{f}. We observe that the exclusion proportions increase with $|\theta|$ for the bounds under Figures \ref{b}, \ref{d}, \ref{f} and \ref{h} (available IV), but are always 0 for the bounds under Figures \ref{a}, \ref{c}, \ref{e} and \ref{f} (no available IV). Hence, with respect to the ability to exclude the causal null hypothesis, it seems preferable to have an available IV than to have an ideal sampling design.

\begin{figure}[ht]
\begin{center}		
\includegraphics[width=.98\textwidth]{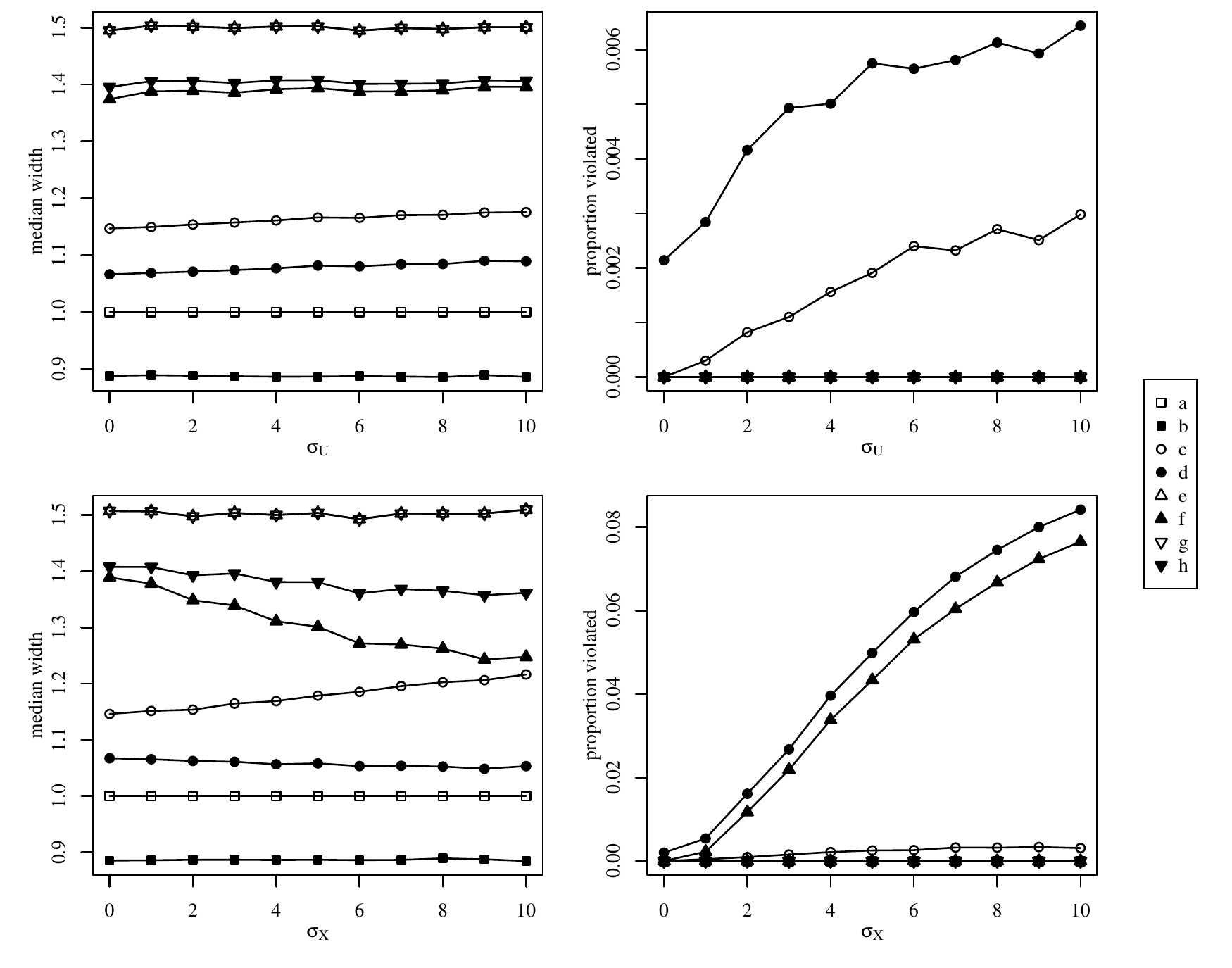}
\caption{Median width of the bounds (left column) and proportion violated (right column), as a function of $\sigma_U$ (top row) and $\sigma_X$ (bottom row).}
\label{fig:width}
\end{center}
\end{figure}

We next repeated the simulation for $\sigma_U=0,1,2,\ldots ,10$, holding $\sigma_X$ fixed at 0 (top row of Figure \ref{fig:width}), and for $\sigma_X=0,1,2,\ldots ,10$, holding $\sigma_U$ fixed at 0 (bottom row of Figure \ref{fig:width}). For each combination of $(\sigma_U,\sigma_X)$ we computed the median width of the bounds (left column of Figure \ref{fig:width}) and the proportion of time the bounds were violated, i.e. the proportion of times the true causal risk difference is outside the bounds (right column of Figure \ref{fig:width}). We observe that the median width is virtually constant across $\sigma_U$ and $\sigma_X$ for most of the bounds, but varies markedly in $\sigma_X$ for bounds under Figures \ref{c}, \ref{f} and \ref{h}. When $\sigma_U>0$, we expect the bounds under Figures \ref{c} and \ref{d} to be sometimes violated, since these bounds assume unconfounded sampling, but the data are generated under confounded sampling. Similarly, when $\sigma_X>0$, the bounds under Figures \ref{c}, \ref{d} and \ref{f} may be violated, since these bounds assume sampling that is not directly dependent on the exposure. We observe that, although violations do indeed occur when $\sigma_U>0$, they are very rare; at $\sigma_U=10$ the bounds under Figures \ref{c} and \ref{d} are violated for less than 1\% of all simulated distributions. When $\sigma_X>0$, violations appear to be relatively common for the bounds under Figures \ref{d} and  \ref{f}; at $\sigma_X=10$ these bounds are violated for almost 10\% of all simulated distributions. However, the bounds under Figure \ref{c} appear to be more robust, with a the risk of violation less than 1\% at $\sigma_X=10$.   

\subsection{Estimated bounds}

To assess finite sample inference for estimated bounds we simulated outcome-dependent samples from the conditional distribution $p\{Z,X,Y|S=1\}$ implied by the model in \eqref{eq:simmod}. For this simulation we fixed the parameter values at $p\{U=1\}=p\{Z=1\}=0.5$, $\alpha_1=\beta_1=\gamma_1=-1$, $\alpha_2=\alpha_3=\beta_2=\beta_3=\gamma_2=0.5$, and $\alpha_4=\beta_4=\gamma_3=\gamma_4=0$; by setting $\gamma_3$ and $\gamma_4$ equal to 0 the data are generated under Figure \ref{d} (unconfounded outcome-dependent sampling). For these parameter values the true causal risk difference is 0.12, the selection probability $p(S=1)$ is equal to 0.31, and the true bounds under Figures \ref{c}-\ref{h} are given in Table \ref{tab:simtrue}.

We drew 1000 samples from $p\{Z,X,Y|S=1\}$, each with a fixed sample size $n$, thus mimicking a scenario where $p\{S=1\}=n/N$ is fixed and known. We carried out the simulation with $n=100$ and $n=1000$. From each sample we drew 1000 bootstrap samples of Type A (see Section \ref{est}), and for each bootstrap sample we estimated the bounds under Figures \ref{c}-\ref{h}, using the true value of $p\{S=1\}$ and (for the bounds under Figures \ref{d}, \ref{f} and \ref{h}) $p\{Z=1\}$. We used the (0.025,0.975) quantiles in the bootstrap distributions for the estimated lower and upper bounds as (0.025,0.975) confidence limits for the true lower and upper bounds, respectively. We computed the mean bias and standard deviation (over the 1000 samples) of the estimated bounds, and the coverage probability of the 95\% bootstrap confidence intervals.

\begin{table}[h]
    \centering
    \begin{tabular}{ccccccc}
\hline
   &c&d&e&f&g&h\\
   \hline
      lower &  -0.50& -0.46& -0.82& -0.80& -0.82& -0.80\\
    \hline
    upper &   0.64&  0.60&  0.87&  0.85&  0.87&  0.85\\
    
    \hline
    \end{tabular}
    \caption{True bounds under the data generating model.}
    \label{tab:simtrue}
\end{table}

We next repeated the simulation, now drawing finite random samples from $p\{Z,X,Y,S\}$ with fixed sizes $N$. Under this sampling design the observed (i.e. with $S=1$) sample size $n$ varies, thus mimicking a scenario where $n/N$ is a random estimator of $p\{S=1\}$. We carried out the simulation with $N=323$ and $N=3223$, so that $E(n)=p\{S=1\}N$ was 100 and 1000, respectively. To reflect the uncertainty in the estimated $p\{S=1\}$, we used Type B bootstrap estimation (see Section \ref{est}).  

The mean bias of the estimated bounds was 0 to the second decimal, for both lower and upper bounds, both sample sizes, and both sampling designs. Table \ref{tab:simboot} shows the standard deviation of the estimated bounds, and the coverage probability of the 95\% bootstrap confidence intervals. We observe that the standard deviation is reduced when the sample size is increased, but is not larger (up to second decimal) when $p\{S=1\}$ is estimated than when $p\{S=1\}$ is known. We further observe that a few of the confidence intervals have somewhat too small or too large coverage probability, but most have nearly 95\% coverage.

\begin{table}[h]
    \centering
    \resizebox{\textwidth}{!}{
    \begin{tabular}{cccccccccccccc}
\hline
 &\multicolumn{6}{c}{standard deviation} & \multicolumn{6}{c}{coverage probability}\\
   &c&d&e&f&g&h & &c&d&e&f&g&h\\
   \hline
   $p\{S=1\}$ fixed and known &&&&&&&&&&&&&\\
   \hline
   lower bound &&&&&&&&&&&&&\\
   $n=100$ & 0.06 & 0.07 & 0.02 & 0.03 & 0.02 & 0.02 & & 0.95 & 0.98 & 0.95 & 0.92 & 0.95 & 0.93 \\ 
  $n=1000$ & 0.02 & 0.03 & 0.00 & 0.01 & 0.00 & 0.01 & & 0.97 & 0.96 & 0.97 & 0.96 & 0.97 & 0.96 \\ 
  upper bound &&&&&&&&&&&&&\\
   $n=100$ & 0.06 & 0.07 & 0.02 & 0.02 & 0.02 & 0.02 & & 0.94 & 0.98 & 0.95 & 0.91 & 0.95 & 0.91 \\ 
 $n=1000$ & 0.02 & 0.02 & 0.00 & 0.01 & 0.00 & 0.01 & & 0.96 & 0.96 & 0.97 & 0.96 & 0.97 & 0.96 \\ 
  \hline
  $p\{S=1\}$ estimated &&&&&&&&&&&&&\\
  \hline
  lower bound &&&&&&&&&&&&&\\
  $E(n)=100$ & 0.06 & 0.07 & 0.02 & 0.03 & 0.02 & 0.03 & & 0.95 & 0.98 & 0.95 & 0.92 & 0.95 & 0.92 \\ 
  $E(n)=1000$ & 0.02 & 0.03 & 0.01 & 0.01 & 0.01 & 0.01 & & 0.95 & 0.95 & 0.95 & 0.96 & 0.95 & 0.96 \\  
  upper bound &&&&&&&&&&&&&\\
  $E(n)=100$ & 0.06 & 0.07 & 0.02 & 0.03 & 0.02 & 0.03 & & 0.95 & 0.98 & 0.95 & 0.93 & 0.95 & 0.93 \\ 
  $E(n)=1000$ & 0.02 & 0.02 & 0.01 & 0.01 & 0.01 & 0.01 & & 0.95 & 0.96 & 0.94 & 0.95 & 0.94 & 0.95 \\ 
    
    \hline
    \hline
    \end{tabular}
    }
    \caption{Standard deviation of the estimated bounds and coverage probability of 95\% bootstrap confidence intervals.}
    \label{tab:simboot}
\end{table}

\section{Real Data Example} \label{real}

To illustrate the performance of the bounds we use data from a real cohort study on Vitamin D and mortality, described by \citet{martinussen2019instrumental}. To allow for public availability, the data were slightly mutilated before inclusion in the R package \texttt{ivtools}, whence we obtained the data. The exposure ($X$) is vitamin D level at baseline, measured as serum concentration of 25-OH-D (nmol/L). As vitamin D levels below 30 nmol/L indicate vitamin D deficiency \citep{martinussen2019instrumental}, we used this level as a cutoff for defining a binary exposure. The outcome ($Y$) is death during follow-up. The data also contain an IV, a binary indicator of whether the subject has mutations in the filaggrin gene. \citet{martinussen2019instrumental} used this IV to estimate the causal effect of vitamin D on mortality, assuming a parametric structural Cox model. In contrast, our bounds do not make any parametric model assumptions.

The data constitute a random sample, which makes it possible to compute the bounds corresponding to Figures \ref{a} and \ref{b}; these are given by (-0.74, 0.26) and (-0.71, 0.15), respectively. Thus, for these data, the presence of an IV reduces somewhat the range of possible values for $\theta$. Since the bounds span 0, we cannot rule out a lack of an effect of vitamin D deficiency on death, and the data are consistent with null, small positive effects, and small to moderate negative effects. The bounds do rule out a range of large negative effects and moderate to large positive effects.  

To illustrate the impact of outcome-dependent sampling, we generated a selection variable ($S$) randomly for each subject from a Bernoulli distribution with probability $p\{S=1|Y=y\}=\mbox{expit}(\alpha+\beta y)$. Sampling is therefore dependent on the outcome, but is unconfounded and does not depend on the exposure. We set $\beta$ to 0.5, to simulate sampling that is moderately outcome-dependent. We used three values of $\alpha$, corresponding to the marginal selection probabilities $p\{S=1\} = (0.1, 0.5, 0.9)$. 

For each level of sampling, we estimated the bounds under the assumed causal diagrams in Figures \ref{c}-\ref{h}, using the true values of $p\{S=1\}$ and (for the bounds under Figures \ref{d}, \ref{f} and \ref{h}) $p\{Z=1\}$, together with 95\% Type A bootstrap confidence intervals. Table \ref{realbounds} shows the results. As in the simulation, we observe that an available IV generally makes the bounds narrower, and that stronger deviation from the ideal randomized trial with random sampling generally makes the bounds wider. As expected, we also observe that all bounds become narrower as the probability of selection increases.     

\begin{table}[h]
    \centering
    \resizebox{\textwidth}{!}{
    \begin{tabular}{ccccccc}
\hline
   &c&d&e&f&g&h \\
   \hline
    lower bound&&&&&&\\
  p\{S=1\}=0.1 & -0.90 (-0.93,-0.86) & -0.90 (-0.93,-0.85) & -0.96 (-0.97,-0.96) & -0.96 (-0.97,-0.92) & -0.96 (-0.97,-0.96) & -0.96 (-0.97,-0.94) \\ 
  p\{S=1\}=0.5 & -0.82 (-0.84,-0.81) & -0.82 (-0.83,-0.79) & -0.85 (-0.87,-0.84) & -0.85 (-0.86,-0.80) & -0.85 (-0.87,-0.84) & -0.85 (-0.86,-0.82) \\ 
  p\{S=1\}=0.9 & -0.76 (-0.77,-0.74) & -0.72 (-0.76,-0.66) & -0.76 (-0.78,-0.75) & -0.72 (-0.77,-0.65) & -0.76 (-0.78,-0.75) & -0.76 (-0.78,-0.73) \\ 
  \hline
   upper bound&&&&&&\\
  p\{S=1\}=0.1 & 0.89 (0.85,0.93) & 0.77 (0.43,0.82) & 0.94 (0.93,0.94) & 0.85 (0.75,0.93) & 0.94 (0.93,0.94) & 0.89 (0.84,0.93) \\ 
  p\{S=1\}=0.5 & 0.60 (0.58,0.62) & 0.55 (0.40,0.61) & 0.65 (0.63,0.66) & 0.56 (0.38,0.64) & 0.65 (0.63,0.66) & 0.59 (0.50,0.65) \\ 
  p\{S=1\}=0.9 & 0.33 (0.31,0.34) & 0.27 (0.09,0.33) & 0.34 (0.32,0.35) & 0.27 (0.09,0.34) & 0.34 (0.32,0.35) & 0.28 (0.19,0.35) \\ 
    \hline
    \hline
    \end{tabular}
    }
    \caption{Estimated bounds with 95\% bootstrap confidence intervals under outcome-dependent sampling of the vitamin D data.}
    \label{realbounds}
\end{table}

\section{Discussion} \label{dis}
We have derived nonparametric bounds for the causal risk difference under six causal diagrams, which  cover a large number of outcome-dependent sampling designs. To our knowledge, five of these causal diagrams (Figures \ref{d}-\ref{h}) have not been previously addressed in the literature on bounds. This provides a nearly complete set of bounds for outcome-dependent sampling designs with binary exposure and outcome. In settings with non-binary categorical exposure or outcome, one could use the bounds that we have derived for comparing any two levels or any grouping of levels of the outcome and exposure provided the IV is still binary. When the IV is non-binary categorical, the additional information captured by the larger dimension of the IV may result in tighter bounds \citep{palmer2011nonparametric}; this is a future area of research being investigated by the authors. When any of the variables are continuous, we conjecture that bounds cannot be computed without further parametric assumptions. 

We have assumed throughout that the analyst has sufficient prior knowledge to determine the underlying causal diagram that generated the data; however, this may not always be the case. It may be possible to use the observed data to deduce the presence or absence of causal relationships, e.g., by using causal discovery algorithms such as the Fast Causal Inference (FCI) algorithm, which allows for unmeasured confounding and selection \citep{spirtes2000causation, zhang2008completeness}. 

Although our derivations rely, in one way or another, on the linear programming method of \citet{balke1994counterfactual}, modifications and/or extensions to this method were required in all settings, as none of the settings clearly define linear programming problems. In particular, we had to modify the method to allow for the sampling variable $S=1$ to define the observed data distribution, which was not considered in \citet{balke1994counterfactual}, or, to our knowledge, in subsequent literature. Our approaches also provide possible routes for the derivation of bounds in other settings where the constraints are nonlinear due the lack of unmeasured confounding or lack of complete observation, or both, for one or more variables.

In outcome-dependent sampling designs, the most common estimand is the odds ratio. This is because when sampling is uncounfounded (Figures \ref{c} and \ref{d}), the observed odds ratio is collapsible over $S$ and is therefore equal to the population (i.e. marginal over $S$) odds ratio \citep{didelez10}. However, in settings with unmeasured confounding of $X$ and $Y$, the observed odds ratio does not have a causal interpretation. In addition, when the sampling is confounded (Figures \ref{e}-\ref{h}), the observed odds ratio is no longer collapsible over $S$ and thus, not equal to the population odds ratio. Therefore, the odds ratio is not a more natural target than any other causal estimand in the settings that we have considered, and it is, in our opinion, less interpretable than other estimands. 

A setting we did not consider is uncounfounded exposure- and outcome-dependent sampling. Although this is a possible scenario, we do not see this as a likely scenario because a practitioner would have to intentionally randomly sample conditional on the joint distribution of $X, Y$. If that were possible, that would imply the existence of a sampling frame in which one had knowledge of the joint distribution of the outcome and exposure, thereby negating the need to do the study at all in the absence of an IV. One situation that may be plausible is when it is desired to measure one or more potential IVs but it is costly to do so, such as a high dimensional IV setting. In that case, one may consider sampling conditional on $(X, Y)$ and measuring $Z$, however, this is not something we have seen suggested in practice. We also do not discuss scenarios where the exposure is randomized, but the sampling is outcome-dependent and/or confounded with outcome. Although this is an interesting set of problems, this is beyond the scope of this paper and an area of future research for the authors. 

\subsubsection*{Supplementary Materials}
These include details on the derivations, and R code to reproduce the numerical results. 

\subsubsection*{Acknowledgments}
We are grateful to the associate editor and two anonymous reviewers for their comments that have improved the article.


\bibliographystyle{abbrvnat}
\bibliography{tndref}
\end{document}


\def\spacingset#1{\renewcommand{\baselinestretch}%
{#1}\small\normalsize} \spacingset{1.5}


\if1\blind
{
  \title{\bf Supplementary material for ``Causal bounds for outcome-dependent sampling in observational studies''}
  \author{Erin E. Gabriel$^{1\star}$\thanks{EEG is partially supported by Swedish Research Council grant 2017-01898, AS by Swedish Research Council grant 2016-01267 and MCS by Swedish Research Council grant 2019-00227} \hspace{.2cm}\\
Michael C. Sachs$^1$\\
Arvid Sj\"olander$^{1\star}$\\
\footnotesize 1:Department of Medical Epidemiology and Biostatistics, Karolinska Institutet, Stockholm, Sweden \\
$\star$The authors contributed equally to this work.\\
corresponding author: erin.gabriel@ki.se}
  \maketitle
} \fi 

\if0\blind
{
  \bigskip
  \bigskip
  \bigskip
  \begin{center}
    {\LARGE\bf Supplementary material for ``Causal bounds for outcome-dependent sampling in observational studies''}
\end{center}
  \medskip
} \fi

\section{IV bounds for random sampling}
Define $p_{xy.Z=z}=p\{X=x,Y=y|Z=z\}$. \cite{Balke97} showed that 
\begin{eqnarray}
\label{eq:b18}
\theta\geq \max\left\{\begin{array}{l}p_{11.Z=1}+p_{00.Z=0}-1\\
p_{11.Z=0}+p_{00.Z=1}-1\\
p_{11.Z=0}-p_{11.Z=1}-p_{01.Z=1}-p_{10.Z=0}-p_{01.Z=0}\\
p_{11.Z=1}-p_{11.Z=0}-p_{01.Z=0}-p_{10.Z=1}-p_{01.Z=1}\\
-p_{10.Z=1}-p_{01.Z=1}\\
-p_{10.Z=0}-p_{01.Z=0}\\
p_{00.Z=1}-p_{10.Z=1}-p_{01.Z=1}-p_{10.Z=0}-p_{00.Z=0}\\
p_{00.Z=0}-p_{10.Z=0}-p_{01.Z=0}-p_{10.Z=1}-p_{00.Z=1}
\end{array}\right\}
\end{eqnarray}
and 
\begin{eqnarray}
\label{eq:b19}
\theta\leq \min\left\{\begin{array}{l}1-p_{10.Z=1}-p_{01.Z=0}\\
1-p_{10.Z=0}-p_{01.Z=1}\\
-p_{10.Z=0}+p_{10.Z=1}+p_{00.Z=1}+p_{11.Z=0}+p_{00.Z=0}\\
-p_{10.Z=1}+p_{11.Z=1}+p_{00.Z=1}+p_{10.Z=0}+p_{00.Z=0}\\
p_{11.Z=1}+p_{00.Z=1}\\
p_{11.Z=0}+p_{00.Z=0}\\
-p_{01.Z=1}+p_{11.Z=1}+p_{00.Z=1}+p_{11.Z=0}+p_{01.Z=0}\\
-p_{01.Z=0}+p_{11.Z=0}+p_{00.Z=0}+p_{11.Z=1}+p_{01.Z=1}
\end{array}\right\},
\end{eqnarray}
in the setting of Figure 1b of the main text. \

\section{Proof of Result 1: Bounds under unconfounded outcome-dependent sampling with no available IV} \label{sec:prel}

As defined in the main text: 
\begin{itemize}
    \item $p_{xy}=p\{X=x,Y=y\}$ 
    \item $p_{xy.s}=p\{X=x,Y=y|S=s\}$ 
    \item $r=p\{S=1\}$
\end{itemize}

In the setting of Figure 1c of the main text, the data distribution is defined by 8 probabilities conditional on the selection variable $S$ and the unknown probabilities $(p_{00.0},p_{01.0},p_{10.0},p_{11.0})$ are constrained by
\begin{eqnarray}
\label{eq:1}
\sum_{xy}p_{xy.0}=1.
\end{eqnarray} 
If we assume that sampling was not deterministic, i.e. $0<p\{S=1|Y=y\}<1$  and that all observed probabilities are $0<p_{xy.1}<1$ then  
\begin{eqnarray}
\label{eq:Ay}
A(y,0)=A(y,1)\textrm{ for all } y.
\end{eqnarray}
This can be seen because $S\bot X|Y$ and  
\begin{eqnarray*}
&&\frac{p\{X=1|Y=y,S=s\}}{p\{X=0|Y=y,S=s\}}\\
&&\phantom{1}=\frac{p_{1y.s}/p\{Y=y|S=s\}}{p_{0y.s}/p\{Y=y|S=s\}}\\
&&\phantom{1}=A(y,s)
\end{eqnarray*}
does not depend on $s$. 

Starting from the bounds derived by \citet{robins1989analysis} and shown to be tight by \citet{Balke97}, 
\begin{eqnarray}
\label{Beq:1}
-(p_{01}+p_{10}) \leq \theta \leq 1-(p_{01}+p_{10}),
\end{eqnarray}
we can expand and derive bounds in our setting. To derive bounds for $\theta$ in the setting of Figure 1c that take the constraint in \eqref{eq:Ay} into account we proceed as follows. We partition the lower limit term $p_{01}+p_{10}$ in \eqref{Beq:1} to 
\begin{eqnarray}
\label{eq:partitioning}
-(p_{01}+p_{10})=-(p_{01.1}+p_{10.1})r-(p_{01.0}+p_{10.0})(1-r).
\end{eqnarray}
Under \eqref{eq:1} and \eqref{eq:Ay} we can express $(p_{01.0},p_{10.0},p_{11.0})$ as functions of $p_{00.0}$. From \eqref{eq:Ay} we have  
\begin{eqnarray}
\label{eq:4}
p_{10.0}=p_{00.0}A(0,1).
\end{eqnarray}
We further have 
\begin{eqnarray}
\label{eq:5}
p_{01.0}&=&1-p_{00.0}-p_{10.0}-p_{11.0}\nonumber\\
&=&1-p_{00.0}\{1+A(0,1)\}-p_{11.0},
\end{eqnarray}
where the first equality follows from (\ref{eq:1}) and the second from (\ref{eq:4}). Plugging (\ref{eq:5}) into (\ref{eq:Ay}) gives
\begin{eqnarray*}
\frac{p_{11.0}}{1-p_{00.0}\{1+A(0,1)\}-p_{11.0}}=A(1,1),
\end{eqnarray*}
which gives 
\begin{eqnarray}
\label{eq:6}
p_{11.0}=\frac{A(1,1)[1-p_{00.0}\{1+A(0,1)\}]}{1+A(1,1)}.
\end{eqnarray}
Finally, plugging (\ref{eq:6}) into (\ref{eq:5}) gives
\begin{eqnarray}
\label{eq:7}
p_{01.0}=\frac{1-p_{00.0}\{1+A(0,1)\}}{1+A(1,1)}.
\end{eqnarray}
We proceed by finding the feasible space for $p_{00.0}$, i.e. the space for which $(p_{00.0},p_{01.0},p_{10.0},p_{11.0})$ are all $\in [0,1]$. $(p_{01.0},p_{10.0},p_{11.0})$ are monotone functions of $p_{00.0}$. Thus, the feasible space for $p_{00.0}$ is a continuous range with $p_{00.0} > 0$. To find the upper bound for $p_{00.0}$, note that $p_{10.0}$ is increasing in $p_{00.0}$, with $p_{10.0}$ approaching 1 when $p_{00.0}$ approaches $1/A(0,1)$; $p_{01.0}$ is decreasing in $p_{00.0}$, with $p_{01.0}$ approaching zero when $p_{00.0}$ approaches $1/\{1+A(0,1)\}$; $p_{11.0}$ is decreasing in $p_{00. 0}$, with $p_{11.0}$ approaching zero when $p_{00.0}$ approaches $1/\{1+A(0,1)\}$. Thus,  $p_{00.0}\leq \min[1/A(0,1),1/\{1+A(0,1)\}]=1/\{1+A(0,1)\}$. We thus have that 
\begin{eqnarray}
\label{eq:8}
0\leq p_{00.0} \leq \frac{1}{1+A(0,1)}.
\end{eqnarray}
From (\ref{eq:4}) and (\ref{eq:7}) we have that 
\begin{eqnarray}
\label{eq:9}
p_{01.0}+p_{10.0}=\frac{1+p_{00.0}[A(0,1)\{1+A(1,1)\}-\{1+A(0,1)\}]}{1+A(1,1)}.
\end{eqnarray}
If $A(0,1)/\{1+A(0,1)\}\geq 1/\{1+A(1,1)\}$, then the right hand side of (\ref{eq:9}) is increasing in $p_{00.0}$. Using (\ref{eq:8}) we thus have that 
\begin{eqnarray}
\label{eq:14}
\min(p_{01.0}+p_{10.0})&=&\frac{1}{1+A(1,1)},\nonumber\\
\max(p_{01.0}+p_{10.0})&=&\frac{A(0,1)}{1+A(0,1)}.
\end{eqnarray} 
If $A(0,1)/\{1+A(0,1)\}\leq 1/\{1+A(1,1)\}$, then the right hand side of (\ref{eq:9}) is decreasing in $p_{00.0}$. Using (\ref{eq:8}) we thus have that 
\begin{eqnarray}
\label{eq:15}
\min(p_{01.0}+p_{10.0})&=&\frac{A(0,1)}{1+A(0,1)},\nonumber\\
\max(p_{01.0}+p_{10.0})&=&\frac{1}{1+A(1,1)}.
\end{eqnarray}
Combining (\ref{eq:14}) and (\ref{eq:15}) gives
\begin{eqnarray}
\label{eq:16}
\min(p_{01.0}+p_{10.0})&=&\min\left\{\frac{1}{1+A(1,1)},\frac{A(0,1)}{1+A(0,1)}\right\},\nonumber\\
\max(p_{01.0}+p_{10.0})&=&\max\left\{\frac{1}{1+A(1,1)},\frac{A(0,1)}{1+A(0,1)}\right\}.
\end{eqnarray}
Combining (\ref{eq:partitioning}) and (\ref{eq:16}) gives
\begin{eqnarray}
\label{eq:17}
\theta\geq -(p_{01.1}+p_{10.1})r-\max\left\{\frac{1}{1+A(1,1)},\frac{A(0,1)}{1+A(0,1)}\right\}(1-r)
\end{eqnarray}
and
\begin{eqnarray}
\label{eq:18}
\theta\leq 1-(p_{01.1}+p_{10.1})r-\min\left\{\frac{1}{1+A(1,1)},\frac{A(0,1)}{1+A(0,1)}\right\}(1-r).
\end{eqnarray}

Therefore the bounds given in the main text for $\theta$ are valid in the setting of Figure 1c, provided that sampling was not deterministic, i.e. $0<p\{S=1|Y=y\}<1$  and that all observed probabilities are $0<p_{xy.1}<1$. We consider the bounds when this is not true on a case by case basis. 

\vspace{3mm}

\noindent \textbf{Case 0}

Let:
\begin{itemize}
\item $0<p\{S=1|Y=y\}<1$ $\forall y \in \{0,1\}$, 
\item $0<p_{00.1}<1$ and $0<p_{01.1}<1$
\item $p_{10.1}=0$ or $p_{11.1}=0$ or both. 
\end{itemize}

Then the constraints still hold as above:
\begin{eqnarray*}
\sum_{xy}p_{xy.0}&=&1\\
\frac{p_{10.0}}{p_{00. 0}}&=&A(0,1)\\
\frac{p_{11. 0}}{p_{01. 0}}&=&A(1,1).
\end{eqnarray*}
In this setting, $A(0,1)$ or $A(1,1)$ or both are equal to 0. As the values of $A(0,1)$ and $A(1,1)$, given that they are defined, are never considered in the derivations, this will not change the resulting bounds. All derivations from above follow in this case. We will call this set of bounds the primary bounds. 
\vspace{1cm}

\noindent \textbf{Case 1}

Let:
\begin{itemize}
    \item $0<p\{S=1|Y=y\}<1$ $\forall y \in \{0,1\}$
    \item $0<p_{10.1}<1$ and $0<p_{11.1}<1$
    \item $p_{00.1}=0$ or $p_{01.1}=0$ or both. 
\end{itemize}

Here, $A(0,1)$ or $A(1,1)$ or both are undefined, but $A(0,1)^{-1}$ or $A(1,1)^{-1}$ are defined. Therefore, we can switch the meaning of exposure, relabeling $X$ to 1 for unexposed and 0 for exposed, putting us back in the primary case setting, and the bounds derivations hold as above for $\theta^* = -\theta$, with the upper $u^*$ and lower $l^*$ bounds derived in this way giving the bounds $-u^*\leq \theta \leq -l^*$ for $\theta$.
\vspace{1cm}

\noindent \textbf{Case 2}

Let:
\begin{itemize}
    \item $0<p\{S=1|Y=y\}<1$ $\forall y \in \{0,1\}$
    \item $0<p_{01.1}<1$ and $0<p_{11.1}<1$
    \item $p_{00.1}=0$ and $p_{10.1}=0$
\end{itemize}

Here $A(0,1)$ and $A(0,1)^{-1}$ are both undefined, but if $0<p\{S=1|Y=y\}<1$ $\forall y \in \{0,1\}$, then $p_{xy.1}=0$ if and only if $p_{xy}=0$.\\

\noindent This is because 
$$p_{xy.s} = \frac{p\{s|x,y\}*p\{x,y\}}{p\{s\}},$$ 

\noindent and $p\{s|x,y\} =p\{s|y\}$ here because $S\bot X|Y$. Therefore, if $0<p\{S=1|Y=y\}<1$, $p_{xy.1}=0$ iff $p_{xy}=0$. Thus, $p_{00.1}=0$ and $p_{10.1}=0$ implies that $p_{00.0}=0$ and $p_{10.0}=0$.\\

\noindent There are now two equations and two unknowns, $p_{11. 0}/p_{01. 0}=A(1,1)$ and $p_{11.0}+p_{01.0}=1$, we can use these to solve for the value of $$p_{01. 0}= \frac{1}{1+A(1,1)}.$$\\

\noindent Going back to the original lower bound in equation \eqref{eq:partitioning}, 
$$-(p_{01.1}+p_{10.1})r-(p_{01.0}+p_{10.0})(1-r)$$
and replacing $p_{01. 0}$ in the lower bound with $\frac{1}{1+A(1,1)}$ and setting $p_{10.1}=p_{10.0}=0$, we get 
$$-(p_{01.1})r-\left(\frac{1}{1+A(1,1)}\right)(1-r).$$ 
This can also be applied to the upper bound to obtain
$$1-(p_{01.1})r-\left(\frac{1}{1+A(1,1)}\right)(1-r).$$ 
This will be the same bound that would be obtained by removing the $\frac{A(0,1)}{1+A(0,1)}$ from the maximum and minimum terms from the bounds in Equations \eqref{eq:17} and \eqref{eq:18}, setting them to $\frac{1}{1+A(1,1)}$.\\

Similarly, if $A(1,1)$ and $A(1,1)^{-1}$ are both undefined, then removing $\frac{1}{1+A(1,1)}$ from the bounds sets them to
$$-(p_{10.1})r-\left(\frac{A(0,1)}{1+A(0,1)}\right)(1-r),$$ 
and 
$$1-(p_{10.1})r-\left(\frac{A(0,1)}{1+A(0,1)}\right)(1-r).$$
This follows by noting that the max and min of $p_{10.0}$ are both $\frac{A(0,1)}{1+A(0,1)}$ because we know that $p_{01.0}=0$. This covers the case where 
$0<p_{00.1}<1$ and $0<p_{10.1}<1$ and $p_{11.1}=0$ and $p_{01.1}=0$.

\vspace{1cm}

\noindent \textbf{Case 3}

Let:
\begin{itemize}
    \item $P\{S=1|Y=1\}=1$ and $0<P\{S=1|Y=0\}<1$
    \item $0<p_{01.1}<1$ and $0<p_{00.1}<1$ 
\end{itemize}

Here, neither $A(1,1)$ nor $A(0,1)$ are undefined, but $p_{11.0}=p_{01.0}=0$ and therefore their ratio is not equal to $A(1,1)$. Thus, we have the following constraints:
\begin{eqnarray*}
\sum_{xy}p_{xy.0}&=&1\\
A(0, 1)=\frac{p_{10.1}}{p_{00. 1}}&=&\frac{p_{10.0}}{p_{00. 0}}\\
A(1, 1)=\frac{p_{11. 1}}{p_{01. 1}}&\neq&\frac{p_{11. 0}}{p_{01. 0}}.
\end{eqnarray*}
Given that we know that $p_{11. 0}=p_{01. 0}=0$, we are back in case 2, and $\frac{1}{1+A(1,1)}$ can simply be removed from the bounds. However, if $\frac{1}{1+A(1,1)}>\frac{A(0,1)}{1+A(0,1)}$, then the lower bound will be lower than needed, and the upper bound will be correct and if $\frac{1}{1+A(1,1)}<\frac{A(0,1)}{1+A(0,1)}$ then the upper bounds will be higher than needed, but the lower bound will be correct. Therefore, the bounds for $\theta$ given in Equations \eqref{eq:17} and \eqref{eq:18} are valid in this case.

If $p_{01.1}=0$ or $p_{00.1}=0$ or both, using the same reasoning as in case 1, if $0<p_{11.1}<1$ and $0<p_{01.1}<1$ the bounds given in \eqref{eq:17} and \eqref{eq:18} are valid for $\theta^*$. \\

\noindent This holds similarly if $p\{S=1|Y=0\}=1$ and $0<p\{S=1|Y=1\}<1$. \\

\noindent If both $p\{S=1|Y=1\}=1$ and $p\{S=1|Y=0\}=1$, then there is no sampling and we are back in the original case, which will always have tighter bounds than the stated primary or inverted bounds with sampling. This again makes the primary or inverted bounds valid but not tight. 
\vspace{1cm}

\noindent \textbf{Case 4}

Let:
\begin{itemize}
    \item $p\{S=1|Y=1\}=0$ and $0<p\{S=1|Y=0\}<1$
    \item $p_{00.1}>0$ and $p_{10.1}>0$ 
\end{itemize}

Here, $A(1,1)$ is undefined and $\frac{p_{11. 1}}{p_{01. 1}}\neq\frac{p_{11. 0}}{p_{01. 0}}$, but 
\begin{eqnarray*}
\sum_{xy}p_{xy.0}&=&1\\
A(0,1)=\frac{p_{10.1}}{p_{00. 1}}&=&\frac{p_{10.0}}{p_{00. 0}}.\\
\end{eqnarray*}

This gives us no information about $p_{11. 0}$ or $p_{01. 0}$. Therefore we cannot use the primary bounds or the reasoning from cases 0-3. \\

\noindent The constraints bound $p_{01.0}<1$ when $p_{00.1}>0$ and $p_{10.1}>0$. However, any attempt to further refine this constraint requires information about one of the unknown values, and thus the best we can do is bound with 1. There is nothing constraining $p_{01.0}$ away from zero. \\

\noindent For this reason the only always valid bounds based on the available information are: 
$$-(p_{10.1})r-(1-r),$$ 
and 
$$1-(p_{10.1})r.$$ 
In the case when $p\{S=1|Y=0\}=0$ and $0<p\{S=1|Y=1\}<1$, we can similarly derive the bounds

$$-(p_{01.1})r-(1-r),$$ 
and 
$$1-(p_{01.1})r.$$ 

Combining these two results, we can write the bounds for the case where either $p\{S=1|Y=0\}=0$ or $p\{S=1|Y=1\}=0$ as
$$-(p_{10.1}+p_{01.1})r-(1-r),$$ 
and 
$$1-(p_{10.1}+p_{01.1})r.$$ 
These are equivalent to the confounded bounds given the main text. \\

\noindent In summary, if one knows the conditional sampling probabilities $0<p\{S=1|Y=y\}<1$ for all $y$, then this can allow for the derivation of tighter bounds in settings that do not fall under those considered in the derivation of the primary bounds. If one does not know the conditional sampling probabilities, but either $A(1,1)$ and $A(0,1)$ or $A(1,1)^{-1}$ and $A(0,1)^{-1}$ are not undefined, then the bounds given in \eqref{eq:17} and \eqref{eq:18} are valid for $\theta$ or $\theta^*$. When either both $A(1,1)$ and $A(1,1)^{-1}$ or $A(0,1)$ and $A(0,1)^{-1}$ are undefined, and the conditional sampling probability is unknown, we cannot distinguish between case 2 and case 4. For this reason, one must default to the confounded bounds to guarantee validity.\\

Result 1 of the main text is then proven by the derivation of the bounds \eqref{eq:17} and \eqref{eq:18} when  $0<p\{S=1|Y=y\}<1$ for all $y$ and $0<p_{0y.1}<1$, and by the arguments given in cases 0, 1 and 3 when there are deviations from $0<p\{S=1|Y=y\}<1$ for all $y$ and $0<p_{0y.1}<1$ that do not result in both $A(1,1)$ and $A(1,1)^{-1}$ or $A(0,1)$ and $A(0,1)^{-1}$ being undefined. Finally, the argument provided in case 4 shows why the confounded bounds are needed when both $A(1,1)$ and $A(1,1)^{-1}$ or $A(0,1)$ and $A(0,1)^{-1}$ are undefined and the conditional sampling is unknown. Although not stated in Result 1, case 2 provides alternative bounds that are potentially tighter than the confounded bounds when conditional sampling information is available.

\section{Proof of Result 2: Bounds under unconfounded outcome-dependent sampling with an available IV}

We define: 
\begin{itemize}
    \item $p_{xy.zs}=p\{X=x,Y=y|Z=z,S=s\}$ 
    \item $p_{xy.Z=z}=p\{X=x,Y=y|Z=z\}$ 
    \item $r_z=p\{S=1|Z=z\}$
\end{itemize}

In the setting of Figure 1d of the main text there are 16 probabilities, defined by the selection variable $S$. If we assume that sampling was not deterministic, i.e. $0<p\{S=1|Y=y\}<1$  and that all observed probabilities are $0<p_{xy.z1}<1$ then the unknown probabilities are constrained by
\begin{eqnarray}
\label{eq:b1}
\sum_{xy}p_{xy.z0}=1.
\end{eqnarray} 
We further have that $S\bot X|(Z,Y)$, so that 
\begin{eqnarray*}
&&\frac{p\{X=1|Z=z,Y=y,S=s\}}{p\{X=0|Z=z,Y=y,S=s\}}\\
&&\phantom{1}=\frac{p_{1y.zs}/p\{Y=y|Z=z,S=s\}}{p_{0y.zs}/p\{Y=y|Z=z,S=s\}}\\
&&\phantom{1}=B(y,z,s)
\end{eqnarray*}
does not depend on $S$. We thus have the additional constraints
\begin{eqnarray}
\label{eq:b2}
\frac{p_{10. z0}}{p_{00. z0}}&=&B(0,z,1) \mbox{ and}\\
\label{eq:b3}
\frac{p_{11. z0}}{p_{01. z0}}&=&B(1,z,1).
\end{eqnarray}

Under (\ref{eq:b1})-(\ref{eq:b3}) we can express $(p_{01. z0},p_{10. z0},p_{11. z0})$ as functions of $p_{00. z0}$. From (\ref{eq:b2}) we have that  
\begin{eqnarray}
\label{eq:b4}
p_{10. z0}=p_{00. z0}B(0,z,1).
\end{eqnarray}
We further have that 
\begin{eqnarray}
\label{eq:b5}
p_{01.z0}&=&1-p_{00. z0}-p_{10. z0}-p_{11. z0}\nonumber\\
&=&1-p_{00. z0}\{1+B(0,z,1)\}-p_{11. z0},
\end{eqnarray}
where the first equality follows from (\ref{eq:b1}) and the second from (\ref{eq:b4}). Plugging (\ref{eq:b5}) into (\ref{eq:b3}) gives
\begin{eqnarray*}
\frac{p_{11. z0}}{1-p_{00. z0}\{1+B(0,z,1)\}-p_{11. z0}}=B(1,z,1),
\end{eqnarray*}
which gives 
\begin{eqnarray}
\label{eq:b6}
p_{11. z0}=\frac{B(1,z,1)[1-p_{00.z0}\{1+B(0,z,1)\}]}{1+B(1,z,1)}.
\end{eqnarray}
Finally, plugging (\ref{eq:b6}) into (\ref{eq:b5}) gives
\begin{eqnarray}
\label{eq:b7}
p_{01. z0}=\frac{1-p_{00.z0}\{1+B(0,z,1)\}}{1+B(1,z,1)}.
\end{eqnarray}
We proceed by finding the feasible space for $p_{00. z0}$, i.e. the space for which $(p_{00. z0},p_{01. z0}$, $p_{10. z0},p_{11. z0})$ are all $\in (0,1)$. $(p_{01. z0},p_{10. z0},p_{11. z0})$ are monotone functions of $p_{00. z0}$; thus, the feasible space for $p_{00. z0}$ is a continuous range  $p_{00. z0} > 0$. To find the upper bound for $p_{00. z0}$, note that $p_{10.z0}$ is increasing in $p_{00. z0}$, with $p_{10.z0}$ approaching 1 when $p_{00. z0}$ approaches $1/B(0,z,1)$; $p_{01.z0}$ is decreasing in $p_{00. z0}$, with $p_{01.z0}$ approaching 0  when $p_{00. z0}$ approaches $1/\{1+B(0,z,1)\}$; $p_{11.z0}$ is decreasing in $p_{00. z0}$, with $p_{11.z0}$ approaching 0 when $p_{00. z0}$ approaches $1/\{1+B(0,z,1)\}$. Thus, $p_{00. z0}\leq \min[1/B(0,z,1),1/\{1+B(0,z,1)\}]=1/\{1+B(0,z,1)\}$. We thus have that 
\begin{eqnarray}
\label{eq:b8}
0\leq p_{00. z0} \leq \frac{1}{1+B(0,z,1)}.
\end{eqnarray}

We work with each term in (\ref{eq:b18}) and (\ref{eq:b19}) separately, following the structure of the above derivation without an IV.\\

\noindent\textbf{First term in (\ref{eq:b18})}
\begin{eqnarray*}
&&p_{11.Z=1}+p_{00.Z=0}-1\\
&&\phantom{1}=p_{11.11}r_1+p_{00.01}r_0-1+p_{11.10}(1-r_1)+p_{00.00}(1-r_0)\\
&&\phantom{1}=p_{11.11}r_1+p_{00.01}r_0-1+\frac{B(1,1,1)[1-p_{00.10}\{1+B(0,1,1)\}]}{1+B(1,1,1)}(1-r_1)+p_{00.00}(1-r_0)\\
&&\phantom{1}\geq p_{11.11}r_1+p_{00.01}r_0-1.
\end{eqnarray*}

\noindent\textbf{Second term in (\ref{eq:b18})}
\begin{eqnarray*}
&&p_{11.Z=0}+p_{00.Z=0}-1\\
&&\phantom{1}=p_{11.01}r_0+p_{00.11}r_1-1+p_{11.00}(1-r_0)+p_{00.10}(1-r_1)\\
&&\phantom{1}=p_{11.01}r_0+p_{00.11}r_1-1+\frac{B(1,0,1)[1-p_{00.00}\{1+B(0,0,1)\}]}{1+B(1,0,1)}(1-r_0)+p_{00.10}(1-r_1)\\
&&\phantom{1}\geq p_{11.01}r_0+p_{00.11}r_1-1.
\end{eqnarray*}

\noindent\textbf{Third term in (\ref{eq:b18})}
\begin{eqnarray*}
&&p_{11.Z=0}-p_{11.Z=1}-p_{01.Z=1}-p_{10.Z=0}-p_{01.Z=0}\\
&&\phantom{1}=(p_{11.01}-p_{10.01}-p_{01.01})r_0-(p_{11.11}+p_{01.11})r_1\\
&&\phantom{1}+(p_{11.00}-p_{10.00}-p_{01.00})(1-r_0)-(p_{11.10}+p_{01.10})(1-r_1)\\
&&\phantom{1}=(p_{11.01}-p_{10.01}-p_{01.01})r_0-(p_{11.11}+p_{01.11})r_1\\
&&\phantom{1}+\left(\frac{B(1,0,1)[1-p_{00.00}\{1+B(0,0,1)\}]}{1+B(1,0,1)}-p_{00.00}B(0,0,1)-\frac{1-p_{00.00}\{1+B(0,0,1)\}}{1+B(1,0,1)}\right)(1-r_0)\\
&&\phantom{1}-\left(\frac{B(1,1,1)[1-p_{00.10}\{1+B(0,1,1)\}]}{1+B(1,1,1)}+\frac{1-p_{00.10}\{1+B(0,1,1)\}}{1+B(1,1,1)}\right)(1-r_1)\\
&&\phantom{1}\geq (p_{11.01}-p_{10.01}-p_{01.01})r_0-(p_{11.11}+p_{01.11})r_1-\frac{B(0,0,1)}{1+B(0,0,1)}(1-r_0)-(1-r_1).
\end{eqnarray*}

\noindent\textbf{Fourth term in (\ref{eq:b18})}
\begin{eqnarray*}
&&p_{11.Z=1}-p_{11.Z=0}-p_{01.Z=0}-p_{10.Z=1}-p_{01.Z=1}\\
&&\phantom{1}=(p_{11.11}-p_{10.11}-p_{01.11})r_1-(p_{11.01}+p_{01.01})r_0\\
&&\phantom{1}+(p_{11.10}-p_{10.10}-p_{01.10})(1-r_1)-(p_{11.00}+p_{01.00})(1-r_0)\\
&&\phantom{1}=(p_{11.11}-p_{10.11}-p_{01.11})r_1-(p_{11.01}+p_{01.01})r_0\\
&&\phantom{1}+\left(\frac{B(1,1,1)[1-p_{00.10}\{1+B(0,1,1)\}]}{1+B(1,1,1)}-p_{00.10}B(0,1,1)-\frac{1-p_{00.10}\{1+B(0,1,1)\}}{1+B(1,1,1)}\right)\\
&&\phantom{1}\times(1-r_1)\\
&&\phantom{1}-\left(\frac{B(1,0,1)[1-p_{00.00}\{1+B(0,0,1)\}]}{1+B(1,0,1)}+\frac{1-p_{00.00}\{1+B(0,0,1)\}}{1+B(1,0,1)}\right)(1-r_0)\\
&&\phantom{1}\geq (p_{11.11}-p_{10.11}-p_{01.11})r_1-(p_{11.01}+p_{01.01})r_0-\frac{B(0,1,1)}{1+B(0,1,1)}(1-r_1)-(1-r_0).
\end{eqnarray*}

\noindent\textbf{Fifth term in (\ref{eq:b18})}
\begin{eqnarray*}
&&-p_{10.Z=1}-p_{01.Z=1}\\
&&\phantom{1}=-(p_{10.11}+p_{01.11})r_1-(p_{10.10}+p_{01.10})(1-r_1)\\
&&\phantom{1}=-(p_{10.11}+p_{01.11})r_1-\left[p_{00.10}B(0,1,1)+\frac{1-p_{00.10}\{1+B(0,1,1)\}}{1+B(1,1,1)}\right](1-r_1)\\
&&\phantom{1}=-(p_{10.11}+p_{01.11})r_1-\frac{1+p_{00.10}[B(0,1,1)\{1+B(1,1,1)\}-\{1+B(0,1,1)\}]}{1+B(1,1,1)}(1-r_1)\\
&&\phantom{1}\geq -(p_{10.11}+p_{01.11})r_1-\max\left\{\frac{1}{1+B(1,1,1)},\frac{B(0,1,1)}{1+B(0,1,1)}\right\}(1-r_1).
\end{eqnarray*}

\noindent\textbf{Sixth term in (\ref{eq:b18})}
\begin{eqnarray*}
&&-p_{10.Z=0}-p_{01.Z=0}\\
&&\phantom{1}=-(p_{10.01}+p_{01.01})r_0-(p_{10.00}+p_{01.00})(1-r_0)\\
&&\phantom{1}=-(p_{10.01}+p_{01.01})r_0-\left[p_{00.00}B(0,0,1)+\frac{1-p_{00.00}\{1+B(0,0,1)\}}{1+B(1,0,1)}\right](1-r_0)\\
&&\phantom{1}=-(p_{10.01}+p_{01.01})r_0-\frac{1+p_{00.00}[B(0,0,1)\{1+B(1,0,1)\}-\{1+B(0,0,1)\}]}{1+B(1,0,1)}(1-r_0)\\
&&\phantom{1}\geq -(p_{10.01}+p_{01.01})r_0-\max\left\{\frac{1}{1+B(1,0,1)},\frac{B(0,0,1)}{1+B(0,0,1)}\right\}(1-r_0).
\end{eqnarray*}

\noindent\textbf{Seventh term in (\ref{eq:b18})}
\begin{eqnarray*}
&&p_{00.Z=1}-p_{10.Z=1}-p_{01.Z=1}-p_{10.Z=0}-p_{00.Z=0}\\
&&\phantom{1}=(p_{00.11}-p_{10.11}-p_{01.11})r_1-(p_{10.01}+p_{00.01})r_0\\
&&\phantom{1}+(p_{00.10}-p_{10.10}-p_{01.10})(1-r_1)-(p_{10.00}+p_{00.00})(1-r_0)\\
&&\phantom{1}=(p_{00.11}-p_{10.11}-p_{01.11})r_1-(p_{10.01}+p_{00.01})r_0\\
&&\phantom{1}+\left[p_{00.10}-p_{00.10}B(0,1,1)-\frac{1-p_{00.10}\{1+B(0,1,1)\}}{1+B(1,1,1)}\right](1-r_1)-\{p_{00.00}B(0,0,1)+p_{00.00}\}\\
&&\phantom{1}\times(1-r_0)\\
&&\phantom{1}=(p_{00.11}-p_{10.11}-p_{01.11})r_1-(p_{10.01}+p_{00.01})r_0\\
&&\phantom{1}+\frac{p_{00.10}[\{1-B(0,1,1)\}\{1+B(1,1,1)\}+\{1+B(0,1,1)\}]-1}{1+B(1,1,1)}(1-r_1)-\{B(0,0,1)+1\}p_{00.00}\\
&&\phantom{1}\times(1-r_0)\\
&&\phantom{1}\geq (p_{00.11}-p_{10.11}-p_{01.11})r_1-(p_{10.01}+p_{00.01})r_0\\
&&\phantom{1}-\max\left\{\frac{1}{1+B(1,1,1)},-\frac{1-B(0,1,1)}{1+B(0,1,1)}\right\}(1-r_1)-(1-r_0).
\end{eqnarray*}

\noindent\textbf{Eighth term in (\ref{eq:b18})}
\begin{eqnarray*}
&&p_{00.Z=0}-p_{10.Z=0}-p_{01.Z=0}-p_{10.Z=1}-p_{00.Z=1}\\
&&\phantom{1}=(p_{00.01}-p_{10.01}-p_{01.01})r_0-(p_{10.11}+p_{00.11})r_1\\
&&\phantom{1}+(p_{00.00}-p_{10.00}-p_{01.00})(1-r_0)-(p_{10.10}+p_{00.10})(1-r_1)\\
&&\phantom{1}=(p_{00.01}-p_{10.01}-p_{01.01})r_0-(p_{10.11}+p_{00.11})r_1\\
&&\phantom{1}+\left[p_{00.00}-p_{00.00}B(0,0,1)-\frac{1-p_{00.00}\{1+B(0,0,1)\}}{1+B(1,0,1)}\right](1-r_0)-\{p_{00.10}B(0,1,1)+p_{00.10}\}\\
&&\phantom{1}\times(1-r_1)\\
&&\phantom{1}=(p_{00.01}-p_{10.01}-p_{01.01})r_0-(p_{10.11}+p_{00.11})r_1\\
&&\phantom{1}+\frac{p_{00.00}[\{1-B(0,0,1)\}\{1+B(1,0,1)\}+\{1+B(0,0,1)\}]-1}{1+B(1,0,1)}(1-r_0)-\{B(0,1,1)+1\}p_{00.10}\\
&&\phantom{1}\times(1-r_1)\\
&&\phantom{1}\geq (p_{00.01}-p_{10.01}-p_{01.01})r_0-(p_{10.11}+p_{00.11})r_1\\
&&\phantom{1}-\max\left\{\frac{1}{1+B(1,0,1)},-\frac{1-B(0,0,1)}{1+B(0,0,1)}\right\}(1-r_0)-(1-r_1).
\end{eqnarray*}
Combining the eight terms in (\ref{eq:b18}) thus gives
\begin{eqnarray}\label{lBCIV}
\theta\geq \max\left\{\begin{array}{l}p_{11.11}r_1+p_{00.01}r_0-1\\
p_{11.01}r_0+p_{00.11}r_1-1\\
(p_{11.01}-p_{10.01}-p_{01.01})r_0-(p_{11.11}+p_{01.11})r_1-\frac{B(0,0,1)}{1+B(0,0,1)}(1-r_0)-(1-r_1)\\
(p_{11.11}-p_{10.11}-p_{01.11})r_1-(p_{11.01}+p_{01.01})r_0-\frac{B(0,1,1)}{1+B(0,1,1)}(1-r_1)-(1-r_0)\\
-(p_{10.11}+p_{01.11})r_1-\max\left\{\frac{1}{1+B(1,1,1)},\frac{B(0,1,1)}{1+B(0,1,1)}\right\}(1-r_1)\\
 -(p_{10.01}+p_{01.01})r_0-\max\left\{\frac{1}{1+B(1,0,1)},\frac{B(0,0,1)}{1+B(0,0,1)}\right\}(1-r_0)\\
 (p_{00.11}-p_{10.11}-p_{01.11})r_1-(p_{10.01}+p_{00.01})r_0\\
\phantom{11111}-\max\left\{\frac{1}{1+B(1,1,1)},-\frac{1-B(0,1,1)}{1+B(0,1,1)}\right\}(1-r_1)-(1-r_0)\\
(p_{00.01}-p_{10.01}-p_{01.01})r_0-(p_{10.11}+p_{00.11})r_1\\
\phantom{11111}-\max\left\{\frac{1}{1+B(1,0,1)},-\frac{1-B(0,0,1)}{1+B(0,0,1)}\right\}(1-r_0)-(1-r_1)
\end{array}\right\}.
\end{eqnarray}

\noindent\textbf{First term in (\ref{eq:b19})}
\begin{eqnarray*}
&&1-p_{10.Z=1}-p_{01.Z=0}\\
&&\phantom{1}=1-p_{10.11}r_1-p_{01.01}r_0-p_{10.10}(1-r_1)-p_{01.00}(1-r_0)\\
&&\phantom{1}=1-p_{10.11}r_1-p_{01.01}r_0-p_{00.10}B(0,1,1)-\frac{1-p_{00.00}\{1+B(0,0,1)\}}{1+B(1,0,1)}(1-r_0)\\
&&\phantom{1}\leq 1-p_{10.11}r_1-p_{01.01}r_0.
\end{eqnarray*}

\noindent\textbf{Second term in (\ref{eq:b19})}
\begin{eqnarray*}
&&1-p_{10.Z=0}-p_{01.Z=1}\\
&&\phantom{1}=1-p_{10.01}r_0-p_{01.11}r_1-p_{10.00}(1-r_0)-p_{01.10}(1-r_1)\\
&&\phantom{1}=1-p_{10.01}r_0-p_{01.11}r_1-p_{00.00}B(0,0,1)-\frac{1-p_{00.10}\{1+B(0,1,1)\}}{1+B(1,1,1)}(1-r_1)\\
&&\phantom{1}\leq 1-p_{10.01}r_0-p_{01.11}r_1.
\end{eqnarray*}

\noindent\textbf{Third term in (\ref{eq:b19})}
\begin{eqnarray*}
&&-p_{10.Z=0}+p_{10.Z=1}+p_{00.Z=1}+p_{11.Z=0}+p_{00.Z=0}\\
&&\phantom{1}=(p_{10.11}+p_{00.11})r_1+(p_{11.01}+p_{00.01}-p_{10.01})r_0\\
&&\phantom{1}+(p_{10.10}+p_{00.10})(1-r_1)+(p_{11.00}+p_{00.00}-p_{10.00})(1-r_0)\\
&&\phantom{1}=(p_{10.11}+p_{00.11})r_1+(p_{11.01}+p_{00.01}-p_{10.01})r_0\\
&&\phantom{1}+\{p_{00.10}B(0,1,1)+p_{00.10}\}(1-r_1)\\
&&\phantom{1}+\left(\frac{B(1,0,1)[1-p_{00.00}\{1+B(0,0,1)\}]}{1+B(1,0,1)}+p_{00.00}-p_{00.00}B(0,0,1)\right)(1-r_0)\\
&&\phantom{1}=(p_{10.11}+p_{00.11})r_1+(p_{11.01}+p_{00.01}-p_{10.01})r_0\\
&&\phantom{1}+\{B(0,1,1)+1\}p_{00.10}(1-r_1)\\
&&\phantom{1}+\frac{B(1,0,1)+p_{00.00}[\{1+B(1,0,1)\}\{1-B(0,0,1)\}-B(1,0,1)\{1+B(0,0,1)\}]}{1+B(1,0,1)}(1-r_0)\\
&&\phantom{1}\leq (p_{10.11}+p_{00.11})r_1+(p_{11.01}+p_{00.01}-p_{10.01})r_0\\
&&\phantom{1}+(1-r_1)+\max\left\{\frac{1-B(0,0,1)}{1+B(0,0,1)},\frac{B(1,0,1)}{1+B(1,0,1)}\right\}r_0.
\end{eqnarray*}

\noindent\textbf{Fourth term in (\ref{eq:b19})}
\begin{eqnarray*}
&&-p_{10.Z=1}+p_{11.Z=1}+p_{00.Z=1}+p_{10.Z=0}+p_{00.Z=0}\\
&&\phantom{1}=(p_{11.11}+p_{00.11}-p_{10.11})r_1+(p_{10.01}+p_{00.01})r_0\\
&&\phantom{1}+(p_{11.10}+p_{00.10}-p_{10.10})(1-r_1)+(p_{10.00}+p_{00.00})(1-r_0)\\
&&\phantom{1}=(p_{11.11}+p_{00.11}-p_{10.11})r_1+(p_{10.01}+p_{00.01})r_0\\
&&\phantom{1}+\left(\frac{B(1,1,1)[1-p_{00.10}\{1+B(0,1,1)\}]}{1+B(1,1,1)}+p_{00.10}-p_{00.10}B(0,1,1)\right)(1-r_1)\\
&&\phantom{1}+\{p_{00.00}B(0,0,1)+p_{00.00}\}(1-r_0)\\
&&\phantom{1}=(p_{11.11}+p_{00.11}-p_{10.11})r_1+(p_{10.01}+p_{00.01})r_0\\
&&\phantom{1}+\frac{B(1,1,1)+p_{00.10}[\{1+B(1,1,1)\}\{1-B(0,1,1)\}-B(1.1,1)\{1+B(0,1,1)\}]}{1+B(1,1,1)}(1-r_1)\\
&&\phantom{1}+\{B(0,0,1)+1\}p_{00.00}(1-r_0)\\
&&\phantom{1}\leq (p_{11.11}+p_{00.11}-p_{10.11})r_1+(p_{10.01}+p_{00.01})r_0+\max\left\{\frac{1-B(0,1,1)}{1+B(0,1,1)},\frac{B(1,1,1)}{B(1,1,1)}\right\}(1-r_1)\\
&&\phantom{1}+(1-r_0).
\end{eqnarray*}

\noindent\textbf{Fifth term in (\ref{eq:b19})}
\begin{eqnarray*}
&&p_{11.Z=1}+p_{00.Z=1}\\
&&\phantom{1}=(p_{11.11}+p_{00.11})r_1+(p_{11.10}+p_{00.10})(1-r_1)\\
&&\phantom{1}=(p_{11.11}+p_{00.11})r_1+\left(\frac{B(1,1,1)[1-p_{00.10}\{1+B(0,1,1)\}]}{1+B(1,1,1)}+p_{00.10}\right)(1-r_1)\\
&&\phantom{1}=(p_{11.11}+p_{00.11})r_1+\frac{B(1,1,1)+p_{00.10}[\{1+B(1,1,1)\}-B(1,1,1)\{1+B(0,1,1)\}]}{1+B(1,1,1)}(1-r_1)\\
&&\phantom{1}\leq (p_{11.11}+p_{00.11})r_1+\max\left\{\frac{1}{1+B(0,1,1)},\frac{B(1,1,1)}{1+B(1,1,1)}\right\}(1-r_1).
\end{eqnarray*}

\noindent\textbf{Sixth term in (\ref{eq:b19})}
\begin{eqnarray*}
&&p_{11.Z=0}+p_{00.Z=0}\\
&&\phantom{1}=(p_{11.01}+p_{00.01})r_0+(p_{11.00}+p_{00.00})(1-r_0)\\
&&\phantom{1}=(p_{11.01}+p_{00.01})r_0+\left(\frac{B(1,0,1)[1-p_{00.00}\{1+B(0,0,1)\}]}{1+B(1,0,1)}+p_{00.00}\right)(1-r_0)\\
&&\phantom{1}=(p_{11.01}+p_{00.01})r_0+\frac{B(1,0,1)+p_{00.00}[\{1+B(1,0,1)\}-B(1,0,1)\{1+B(0,0,1)\}]}{1+B(1,0,1)}(1-r_0)\\
&&\phantom{1}\leq (p_{11.01}+p_{00.01})r_0+\max\left\{\frac{1}{1+B(0,0,1)},\frac{B(1,0,1)}{1+B(1,0,1)}\right\}(1-r_0).
\end{eqnarray*}

\noindent\textbf{Seventh term in (\ref{eq:b19})}
\begin{eqnarray*}
&&-p_{01.Z=1}+p_{11.Z=1}+p_{00.Z=1}+p_{11.Z=0}+p_{01.Z=0}\\
&&\phantom{1}=(p_{11.11}+p_{00.11}-p_{01.11})r_1+(p_{11.01}+p_{01.01})r_0\\
&&\phantom{1}+(p_{11.10}+p_{00.10}-p_{01.10})(1-r_1)+(p_{11.00}+p_{01.00})(1-r_0)\\
&&\phantom{1}=(p_{11.11}+p_{00.11}-p_{01.11})r_1+(p_{11.01}+p_{01.01})r_0\\
&&\phantom{1}+\left(\frac{B(1,1,1)[1-p_{00.10}\{1+B(0,1,1)\}]}{1+B(1,1,1)}+p_{00.10}-\frac{1-p_{00.10}\{1+B(0,1,1)\}}{1+B(1,1,1)}\right)(1-r_1)\\
&&\phantom{1}+\left(\frac{B(1,0,1)[1-p_{00.00}\{1+B(0,0,1)\}]}{1+B(1,0,1)}+\frac{1-p_{00.00}\{1+B(0,0,1)\}}{1+B(1,0,1)}\right)(1-r_0)\\
&&\phantom{1}=(p_{11.11}+p_{00.11}-p_{01.11})r_1+(p_{11.01}+p_{01.01})r_0\\
&&\phantom{1}+\frac{-\{1-B(1,1,1)\}+p_{00.10}[\{1+B(0,1,1)\}\{1-B(1,1,1)\}+\{1+B(1,1,1)\}]}{1+B(1,1,1)}(1-r_1)\\
&&\phantom{1}+\frac{\{1+B(1,0,1)\}-p_{00.00}\{1+B(0,0,1)\}\{1+B(1,0,1)\}]}{1+B(1,0,1)}(1-r_0)\\
&&\phantom{1}\leq (p_{11.11}+p_{00.11}-p_{01.11})r_1+(p_{11.01}+p_{01.01})r_0\\
&&\phantom{1}+\max\left\{\frac{1}{B(0,1,1)},-\frac{1-B(1,1,1)}{1+B(1,1,1)}\right\}(1-r_1)+(1-r_0).
\end{eqnarray*}

\noindent\textbf{Eighth term in (\ref{eq:b19})}
\begin{eqnarray*}
&&-p_{01.Z=0}+p_{11.Z=0}+p_{00.Z=0}+p_{11.Z=1}+p_{01.Z=1}\\
&&\phantom{1}=(p_{11.01}+p_{00.01}-p_{01.01})r_0+(p_{11.11}+p_{01.11})r_1\\
&&\phantom{1}+(p_{11.00}+p_{00.00}-p_{01.00})(1-r_0)+(p_{11.10}+p_{01.10})(1-r_1)\\
&&\phantom{1}=(p_{11.01}+p_{00.01}-p_{01.01})r_0+(p_{11.11}+p_{01.11})r_1\\
&&\phantom{1}+\left(\frac{B(1,0,1)[1-p_{00.00}\{1+B(0,0,1)\}]}{1+B(1,0,1)}+p_{00.00}-\frac{1-p_{00.00}\{1+B(0,0,1)\}}{1+B(1,0,1)}\right)(1-r_0)\\
&&\phantom{1}+\left(\frac{B(1,1,1)[1-p_{00.10}\{1+B(0,1,1)\}]}{1+B(1,1,1)}+\frac{1-p_{00.10}\{1+B(0,1,1)\}}{1+B(1,1,1)}\right)(1-r_1)\\
&&\phantom{1}=(p_{11.01}+p_{00.01}-p_{01.01})r_0+(p_{11.11}+p_{01.11})r_1\\
&&\phantom{1}+\frac{-\{1-B(1,0,1)\}+p_{00.00}[\{1+B(0,0,1)\}\{1-B(1,0,1)\}+\{1+B(1,0,1)\}]}{1+B(1,0,1)}(1-r_0)\\
&&\phantom{1}+\frac{\{1+B(1,1,1)\}-p_{00.10}\{1+B(0,1,1)\}\{1+B(1,1,1)\}]}{1+B(1,1,1)}(1-r_1)\\
&&\phantom{1}\leq (p_{11.01}+p_{00.01}-p_{01.01})r_0+(p_{11.11}+p_{01.11})r_1\\
&&\phantom{1}+\max\left\{\frac{1}{B(0,0,1)},-\frac{1-B(1,0,1)}{1+B(1,0,1)}\right\}(1-r_0)+(1-r_1).
\end{eqnarray*}
Combining the eight terms in (\ref{eq:b19}) thus gives
\begin{eqnarray} \label{UBCIV}
\theta\leq \min\left\{\begin{array}{l}1-p_{10.11}r_1-p_{01.01}r_0\\
1-p_{10.01}r_0-p_{01.11}r_1\\
(p_{10.11}+p_{00.11})r_1+(p_{11.01}+p_{00.01}-p_{10.01})r_0\\
\phantom{11111}+(1-r_1)+\max\left\{\frac{1-B(0,0,1)}{1+B(0,0,1)},\frac{B(1,0,1)}{1+B(1,0,1)}\right\}r_0\\
(p_{11.11}+p_{00.11}-p_{10.11})r_1+(p_{10.01}+p_{00.01})r_0\\
\phantom{11111}+\max\left\{\frac{1-B(0,1,1)}{1+B(0,1,1)},\frac{B(1,1,1)}{B(1,1,1)}\right\}(1-r_1)+(1-r_0)\\
(p_{11.11}+p_{00.11})r_1+\max\left\{\frac{1}{1+B(0,1,1)},\frac{B(1,1,1)}{1+B(1,1,1)}\right\}(1-r_1)\\
(p_{11.01}+p_{00.01})r_0+\max\left\{\frac{1}{1+B(0,0,1)},\frac{B(1,0,1)}{1+B(1,0,1)}\right\}(1-r_0)\\
(p_{11.11}+p_{00.11}-p_{01.11})r_1+(p_{11.01}+p_{01.01})r_0\\
\phantom{11111}+\max\left\{\frac{1}{B(0,1,1)},-\frac{1-B(1,1,1)}{1+B(1,1,1)}\right\}(1-r_1)+(1-r_0)\\
(p_{11.01}+p_{00.01}-p_{01.01})r_0+(p_{11.11}+p_{01.11})r_1\\
\phantom{11111}+\max\left\{\frac{1}{B(0,0,1)},-\frac{1-B(1,0,1)}{1+B(1,0,1)}\right\}(1-r_0)+(1-r_1)
\end{array}\right\}.
\end{eqnarray}

\clearpage

\noindent In summary, the bounds given in the main text for $\theta$ are valid in the setting of Figure 1d, provided that sampling was not deterministic, i.e. $0<p\{S=1|Y=y\}<1$  and that all observed probabilities are $0<p_{xy.z1}<1$. When this is not true, the same special cases follow from the setting without IV with minor modification. Following Case 0, 1 and 3 above, if no $B(y,z,s)$ or $B(y,z,s)^{-1}$ are undefined then the bounds given in the primary derivations above, or the $*$ bounds based on the recoding of to $X^*=1-X$, are valid for $\theta$ or $\theta^*=-\theta$, respectively, regardless of the sampling probabilities or the observed probabilities.  

If one knows the conditional sampling probabilities to be $0<p\{S=1|Y=y\}<1$ for all $y$, this provides more information and can allow for the derivation of tighter bounds, even if $B(y,z,s)$ and $B(y,z,s)^{-1}$ are undefined, this follows directly from case 2 above. If, as in case 4, we know the sampling to be deterministic, such that $B(y,z,s)$ and $B(y,z,s)^{-1}$ are undefined for any $y$ and all $z$, then we must instead use the confounded bounds given in the main text for Figure 1f in addition to any remaining terms that are not undefined in the bounds in Equations \eqref{lBCIV} and \eqref{UBCIV}.

If one does not know the conditional sampling probabilities and both  $B(y,z,1)$ and $B(y,z,1)^{-1}$ are undefined for any $y,z$ pair, but not for all levels of $Z$, then this is an additional special case that does not exist in the setting without an IV.\\ 

\noindent \textbf{Case 2$^{\star}$ for IV}\\
Let:
\begin{itemize}
    \item $p\{S=1|Y=y\}$ is unknown 
    \item $B(1,1,1)$ and $B(1,1,1)^{-1}$ are both undefined 
    \item $B(1,0,1)$ and $B(1,0,1)^{-1}$ are not both undefined 
\end{itemize}

Then, unlike in the case without an IV, we know that sampling is not deterministic, i.e. $0<p\{S=1|Y=1\}<1$, because $S \bot Z|Y$. Thus, we are back in the case 2 setting where alternative, tighter bounds can be formed than the confounded bounds. This is because we know that $0<p\{S=1|Y=1\}<1$, and $B(1,1,1)$ and $B(1,1,1)^{-1}$ both being undefined implies that $p_{11.10}=p_{01.10}=0$, for the same reason as is given in case 2 above. This also illustrates why in the setting with an IV, when $B(1,z,1)$ and $B(1,z,1)^{-1}$ are undefined for all $z$, we are in the case 4 setting, without additional information about the sampling.

Result 2 of the main text is then proven by the derivation of the bounds  \eqref{lBCIV} and \eqref{UBCIV}, when  $0<p\{S=1|Y=y\}<1$ for all $y$ and $0<p_{xy.z1}<1$, and by the arguments given in cases 0, 1 and 3 when there are deviations from $0<p\{S=1|Y=y\}<1$ for all $y$ and $0<p_{xy.1}<1$ that do not result in both $B(y,z,s)$ and $B(y,z,s)^{-1}$ being undefined. Finally, the argument provided in case 4 above and case 2$^{\star}$ for IV, here, show why the confounded bounds are needed, in addition to the remaining terms in that are not undefined in  \eqref{lBCIV} and \eqref{UBCIV}, when both $B(y,z,s)$ and $B(y,z,s)^{-1}$ are undefined for any $y$ and all $z$ and the conditional sampling is unknown. Although not stated in Result 2, case 2$^{\star}$ for IV and the extension of case 2 from the setting without an IV provide potentially tighter alternative bounds which will vary with the $B(x,z,1)$ that are undefined. 

\section{Proof of Results 3 and 4} 

Under Figure 1e, the observed data distribution is given by $p\{X,Y|S=1\}$. If we know $p\{S=1\}$, the joint probabilities  $p\{X=x,Y=y,S=1\}=p\{X=x,Y=y|S=1\}p\{S=1\}$ are identifiable. Under Figure 1f the observed data distribution is given by $p\{Z,X,Y|S=1\}$. If we know $p\{S=1\}$ and $p\{Z=1\}$ and therefore, $p\{S=1|Z\}$, the probabilities  $$p\{X=x,Y=y,S=1|Z=z\}=\frac{p\{Z=z,X=x,Y=y|S=1\}}{\sum_{x,y}p\{Z=z,X=x,Y=y|S=1\}}p\{S=1|Z\}$$ are identifiable. These also hold for Figures 1g-1h. The problem from Figure 1e is as follows.\\

\noindent Just as in the main text, let $p_{xy1}=p\{X=x,Y=y,S=1\}$ be the joint probabilities as defined in Table 1. 
\begin{table}[!h]
\centering
\begin{tabular}{lccc}
&$X$ &$Y$& $S$\\
$p_{001}$ &0 &0 &1\\
$p_{101}$ &1 &0& 1\\
$p_{011}$ &0 &1 &1\\
$p_{111}$ &1& 1& 1\\
\end{tabular}
\caption{Observed probabilities in our shorthand notation.}
\end{table}

Let $R_x$, $R_y$, and $R_s$ denote the response function variables which are categorical random variables that define the causal influences of $X$, $Y$, and $S$. The $q$ parameters are probabilities that specify the joint distribution of the response function variables. For more details on this terminology and the algorithm, see \citet{Balke95} and \citet{balke1994counterfactual}.

\begin{table}[!h]
\centering
\begin{tabular}{lccc}
&$R_x$ &$R_y$& $R_s$\\
q000 &0&0&   0 \\ 
 q001 &1&0&   0 \\ 
 q002 &0&1&   0 \\ 
 q003 &1&1&   0 \\ 
 q010 &0&2&   0 \\ 
 q011 &1&2&   0 \\ 
 q012 &0&3&   0 \\ 
 q013 &1&3&   0 \\ 
 q020 &0&0&   1 \\ 
 q021 &1&0&   1 \\ 
 q022 &0&1&   1 \\ 
 q023 &1&1&   1 \\ 
 q030 &0&2&   1 \\ 
 q031 &1&2&   1 \\ 
 q032 &0&3&   1 \\ 
 q033 &1&3&   1 \\ 
 q100 &0&0&   2 \\ 
 q101 &1&0&   2 \\ 
 q102 &0&1&   2 \\ 
 q103 &1&1&   2 \\ 
 q110 &0&2&   2 \\ 
 q111 &1&2&   2 \\ 
 q112 &0&3&   2 \\ 
 q113 &1&3&   2 \\ 
 q120 &0&0&   3 \\ 
 q121 &1&0&   3 \\ 
 q122 &0&1&   3 \\ 
 q123 &1&1&   3 \\ 
 q130 &0&2&   3 \\ 
 q131 &1&2&   3 \\ 
 q132 &0&3&   3 \\ 
 q133 &1&3&   3 \\ 
\end{tabular}
\caption{All possible values of the response function variables.}
\end{table}

\clearpage

The constraints based on the counterfactuals are then given by: 
$$\sum qR_xR_yR_s = 1,$$
and 
\begin{eqnarray*}
p_{001} &=& q001 + q003 + q021 + q023\\ 
p_{101} &=& q101 + q103 + q111 + q113\\          
p_{011} &=& q012 + q013 + q032 + q033\\            
p_{111} &=& q122 + q123 + q132 + q133.    
\end{eqnarray*}

The objective function is the causal risk difference written in terms of the response function variables: 

$$\theta=q020+q021+q022+q023+q120+q121+q122+q12$$ $$-q010-q011-q012-q013-q110-q111-q112-q113.$$
This defines a linear programming problem. The objective function $\theta$ is linear and the constraints are linear. Therefore, the constraints define a convex polytope in the $q$ space. The dual of this problem expresses the edges of this polytope in terms of the observed probabilities, thus by the fundamental theorem of linear programming, the global minimum and maximum of the objective in terms of the observed probabilities can be found by enumerating the vertices of the polytope \citep{DantzigLP}. As the observed probabilities and constraints defined by the linear relationship with the counterfactual $q$ values and the target parameter completely define the bounding problem, this proves the first part of Result 3, the bounds for Figure 1e being valid and tight. The proof for the rest of Result 3 and 4 follows in the same way. 

\section{Proof of Result 5} 

Our proposed bounds for Figure 1c are given by
\begin{eqnarray}
\label{c1}
\theta\geq -(p_{01.1}+p_{10.1})r-\max\left\{\frac{1}{1+A(1,1)},\frac{A(0,1)}{1+A(0,1)}\right\}(1-r)
\end{eqnarray}
\vspace{-0.5cm}
and
\vspace{-0.5cm}
\begin{eqnarray}
\label{c2}
\theta\leq 1-(p_{01.1}+p_{10.1})r-\min\left\{\frac{1}{1+A(1,1)},\frac{A(0,1)}{1+A(0,1)}\right\}(1-r).
\end{eqnarray}
First, note that 
\begin{eqnarray*}
\frac{1}{1+A(1,1)} &=& \frac{p\{X=0,Y=1|S=1\}}{p\{X=0,Y=1|S=1\}+p\{X=1,Y=1|S=1\}}\\&=&\frac{p_{01.1}}{p\{Y=1|S=1\}}
\end{eqnarray*}
and
\begin{eqnarray*}
\frac{A(0,1)}{1+A(0,1)} &=&\frac{p\{X=1,Y=0|S=1\}}{p\{X=1,Y=0|S=1\}+p\{X=0,Y=0|S=1\}}\\ &=&\frac{p_{10.1}}{p\{Y=0|S=1\}}.
\end{eqnarray*}
We can rewrite the bound in \eqref{c1} as 
$$r\left[-(p_{01.1}+p_{10.1})+\max\left\{\frac{1}{1+A(1,1)},\frac{A(0,1)}{1+A(0,1)}\right\}\right] - \max\left\{\frac{1}{1+A(1,1)},\frac{A(0,1)}{1+A(0,1)}\right\},$$
which is monotonically increasing in $r$ if $$ p_{01.1}+p_{10.1}\leq\max\left\{\frac{1}{1+A(1,1)},\frac{A(0,1)}{1+A(0,1)}\right\} .$$
If $\max\left\{\frac{1}{1+A(1,1)},\frac{A(0,1)}{1+A(0,1)}\right\}=\frac{1}{1+A(1,1)}$, then 
$$\frac{p_{10.1}}{p\{Y=0|S=1\}}< \frac{p_{01.1}}{p\{Y=1|S=1\}},$$
which implies that
\begin{eqnarray*}
p_{01.1}+p_{10.1} &<& p_{01.1}\left(1+\frac{p\{Y=0|S=1\}}{p\{Y=1|S=1\}}\right)\\
&=&\frac{p_{01.1}}{p\{Y=1|S=1\}}\\
&=&\frac{1}{1+A(1,1)}.
\end{eqnarray*}
If $\max\left\{\frac{1}{1+A(1,1)},\frac{A(0,1)}{1+A(0,1)}\right\}=\frac{A(0,1)}{1+A(0,1)}$, then 
$$\frac{p_{01.1}}{p\{Y=1|S=1\}}< \frac{p_{10.1}}{p\{Y=0|S=1\}},$$
which implies that
\begin{eqnarray*}
p_{01.1}+p_{10.1} &<& p_{10.1}\left(1+\frac{p\{Y=1|S=1\}}{p\{Y=0|S=1\}}\right)\\
&=&\frac{p_{10.1}}{p\{Y=0|S=1\}}\\
&=&\frac{A(0,1)}{1+A(0,1)}.
\end{eqnarray*}
Hence, the bound in (\ref{c1}) is monotonically increasing in $r$. 

We can rewrite the bound in \eqref{c2} as 
$$1+r\left[-(p_{01.1}+p_{10.1})+\min\left\{\frac{1}{1+A(1,1)},\frac{A(0,1)}{1+A(0,1)}\right\}\right] - \min\left\{\frac{1}{1+A(1,1)},\frac{A(0,1)}{1+A(0,1)}\right\},$$
which is monotonically decreasing in $r$ if $$ p_{01.1}+p_{10.1}\geq\min\left\{\frac{1}{1+A(1,1)},\frac{A(0,1)}{1+A(0,1)}\right\} .$$
If $\min\left\{\frac{1}{1+A(1,1)},\frac{A(0,1)}{1+A(0,1)}\right\}=\frac{1}{1+A(1,1)}$, then 
$$\frac{p_{10.1}}{p\{Y=0|S=1\}}> \frac{p_{01.1}}{p\{Y=1|S=1\}},$$
which implies that
\begin{eqnarray*}
p_{01.1}+p_{10.1} &>& p_{01.1}\left(1+\frac{p\{Y=0|S=1\}}{p\{Y=1|S=1\}}\right)\\
&=&\frac{p_{01.1}}{p\{Y=1|S=1\}}\\
&=&\frac{1}{1+A(1,1)}.
\end{eqnarray*}
If $\min\left\{\frac{1}{1+A(1,1)},\frac{A(0,1)}{1+A(0,1)}\right\}=\frac{A(0,1)}{1+A(0,1)}$, then 
$$\frac{p_{01.1}}{p\{Y=1|S=1\}}> \frac{p_{10.1}}{p\{Y=0|S=1\}},$$
which implies that
\begin{eqnarray*}
p_{01.1}+p_{10.1} &>& p_{10.1}\left(1+\frac{p\{Y=1|S=1\}}{p\{Y=0|S=1\}}\right)\\
&=&\frac{p_{10.1}}{p\{Y=0|S=1\}}\\
&=&\frac{A(0,1)}{1+A(0,1)}.
\end{eqnarray*}
Hence, the bound in (\ref{c2}) is monotonically decreasing in $r$.

Now, note that
\begin{eqnarray*}
\frac{1}{1+A(1,1)} &=&\frac{p_{01.1}}{p\{Y=1|S=1\}}\\
&=&p\{X=0|Y=1,S=1\}\\
&=&p\{X=0|Y=1\}
\end{eqnarray*}
and
\begin{eqnarray*}
\frac{A(0,1)}{1+A(0,1)}&=&\frac{p_{10.1}}{p\{Y=0|S=1\}}\\
&=&p\{X=1|Y=0,S=1\}\\
&=&p\{X=1|Y=0\},
\end{eqnarray*}
where the last equalities follow since $S \perp X|Y$ in the causal diagram in Figure 1c. Thus, when $r$ converges to 0, the bounds in \eqref{c1} and \eqref{c2} converge to
\begin{eqnarray*}
\label{cc1}
-\max\left\{p\{X=0|Y=1\},p\{X=1|Y=0\}\right\}=\min\left\{-p\{X=0|Y=1\},-p\{X=1|Y=0\}\right\}
\end{eqnarray*}
and
\begin{eqnarray*}
\label{cc2}
1-\min\left\{p\{X=0|Y=1\},p\{X=1|Y=0\}\right\}=\max\left\{p\{X=1|Y=1\},p\{X=0|Y=0\}\right\},
\end{eqnarray*}
which are the outcome-conditional bounds given by \citet{kuroki2010sharp} (i.e. equation (11) in their paper).

\bibliographystyle{abbrvnat}
\bibliography{tndref.bib}